\documentclass[aps,showpacs,twocolumn,superscriptaddress,longbibliography]{revtex4-2}

\usepackage{lipsum} 
\usepackage{physics}
\usepackage{amsmath} 
\usepackage{amsfonts} 
\usepackage{amssymb} 
\usepackage{epsfig} 
\usepackage{braket}
\usepackage{graphicx}
\usepackage{bbold} 
\usepackage{hyphenat} 
\usepackage{hyperref}  \usepackage{placeins}
\usepackage[dvipsnames]{xcolor}
\usepackage{array} 
\newcolumntype{C}[1]{>{\centering\arraybackslash}m{#1}}
\usepackage{overpic}

\usepackage[normalem]{ulem}
\usepackage{dcolumn} 
\usepackage{booktabs}
\usepackage{diagbox} 
\usepackage{siunitx} 
\usepackage{tabularx}
\usepackage{float}

\AtBeginDocument{ \heavyrulewidth=.08em \lightrulewidth=.05em \cmidrulewidth=.03em \belowrulesep=.65ex \belowbottomsep=0pt \aboverulesep=.4ex \abovetopsep=0pt \cmidrulesep=\doublerulesep \cmidrulekern=.5em \defaultaddspace=.5em }
\begin{document}

\title{Sideband fingerprint of the Leggett mode in terahertz two-dimensional coherent spectroscopy of multiband superconductors}

\author{Silvia Neri}
\affiliation{ Max Planck Institute for Solid State Research, D-70569, Stuttgart, Germany}
\author{Ryo Shimano}
\affiliation{Department of Physics, University of Tokyo, Hongo, Tokyo 113-0033, Japan}
\affiliation{Cryogenic Research Center, University of Tokyo, Tokyo 113-0032, Japan}
\author{Dirk Manske}
\affiliation{ Max Planck Institute for Solid State Research, D-70569, Stuttgart, Germany}
\author{Naoto Tsuji}
\affiliation{Department of Physics, University of Tokyo, Hongo, Tokyo 113-0033, Japan}
\affiliation{RIKEN Center for Emergent Matter Science (CEMS), Wako, Saitama 351-0198, Japan}
\affiliation{Trans-scale Quantum Science Institute, University of Tokyo, Hongo, Tokyo 113-0033, Japan}
\begin{abstract}
Collective modes of the superconducting condensate encode 
fundamental information on the pairing interaction, yet their direct detection remains experimentally challenging. In multiband superconductors, the Leggett mode, i.e., a coherent oscillation of the relative phase between two condensates, is Raman-active and couples to light only nonlinearly in ordinary situations. Here we show theoretically that terahertz two-dimensional coherent spectroscopy (2DCS) provides a direct fingerprint of the 
Leggett mode (with frequency $\omega_L$) through sideband emission at 
the detection frequency $\omega_t = \omega_L \pm \Omega$ 
($\Omega$ is the central frequency of the incident pulses) in the nonlinear 2D spectrum. 
The key insight is that the finite duration of the THz pulses is the physical mechanism that makes the mode accessible: the broad spectral content of a finite pulse provides the nonlinear frequency combinations needed to resonantly excite the mode at $\omega_L$ without any prior knowledge of its value or tuning of external parameters. We demonstrate robustness of this signature against 
temperature and quasiparticle relaxation, and show that sweeping $\Omega$ allows unambiguous identification of the Leggett mode.
\end{abstract}

\maketitle
\section{Introduction}

\begin{figure}[t] 
    \centering 
 \includegraphics[scale=0.5]{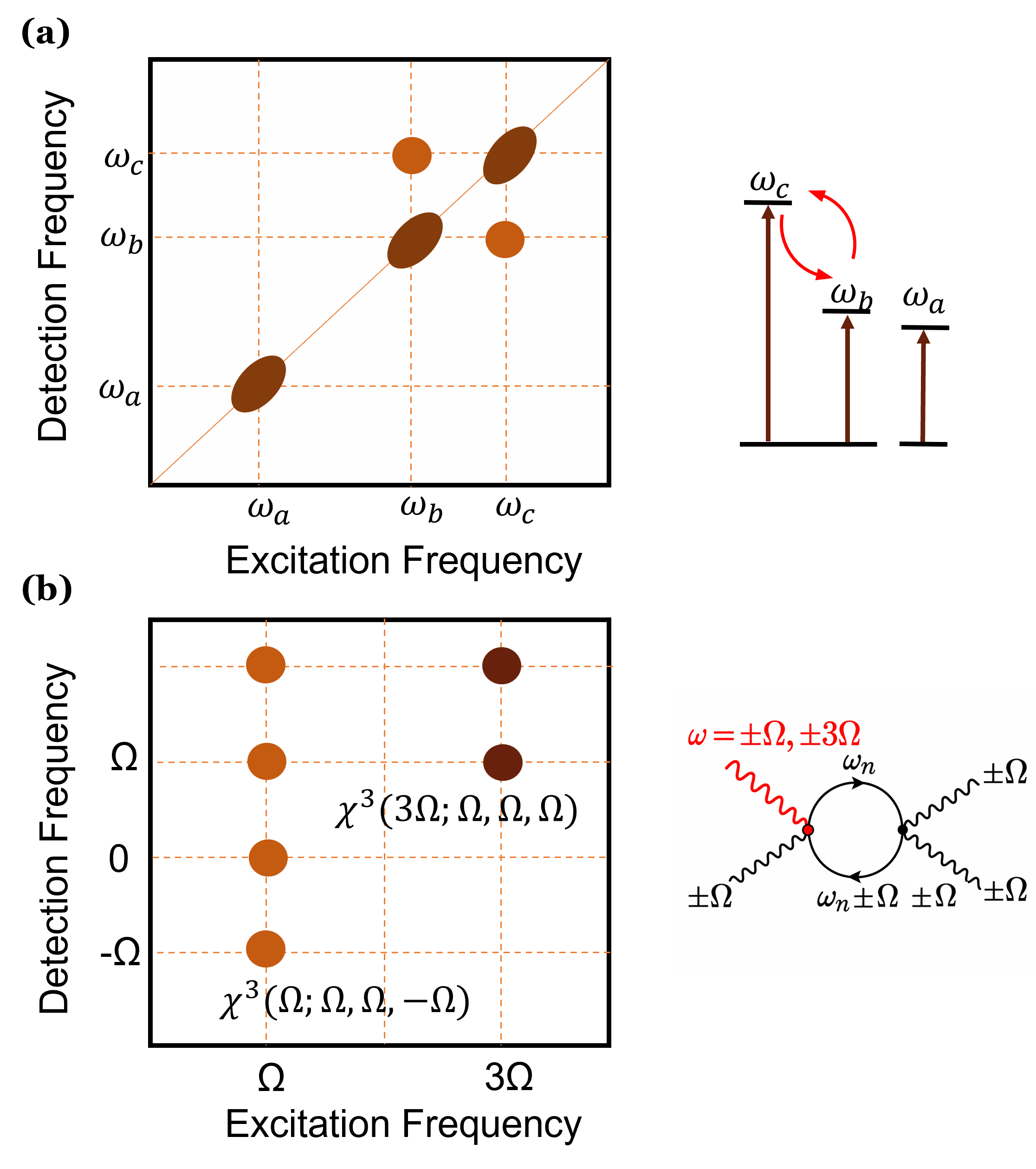} 
    \caption{
    Schematic comparison of (a) conventional 2D spectroscopy originally developed in the field of chemical and molecular physics and (b) 2DCS for quantum materials. (a) Representative 2D spectrum of a mixture of two compounds. Diagonal peaks correspond to conventional 1D spectral features, while off-diagonal peaks reveal coupling between different transitions; peaks B and C belong to the same species and are coupled, as illustrated by the level scheme. (b) Representative 2DCS spectrum generated by two time-delayed pulses, $A(t)$ and $A(t+\Delta t)$, centered at $\pm\Omega$. Third-order frequency mixing produces four fundamental-harmonic spots and two third-harmonic spots. For a clean superconductor, the corresponding third-order diamagnetic process is illustrated on the right, where two incoming photons (wavy lines) couple through diamagnetic vertices to the fermionic propagators (black lines).
    } 
    \label{fig:2d}
\end{figure}

Two-dimensional coherent spectroscopy (2DCS) has emerged as a powerful tool 
for disentangling complex excitation spectra and nonlinear dynamics \cite{Mukamel1995, Fresch2023, Liu2025, Huang2026}.
Originally developed in chemistry to resolve transitions 
associated with molecular vibrations and electronic states,
2DCS has more recently been extended to quantum materials,
providing access to coherent dynamics of
collective excitations such as in quantum magnets \cite{Grishunin2023} and superconductors \cite{Katsumi2024, Katsumi2025, Cheng2025}.
The physical mechanism underlying 2DCS in quantum materials
differs qualitatively from that in conventional molecular spectroscopy.
In molecular systems, 2D signals can often be understood in terms of
transitions between well-defined energy levels connected by distinct
quantum pathways~\cite{Mukamel1995}, as illustrated in Fig.~\ref{fig:2d}(a).
By contrast, in quantum materials 2D signals are generated by 
quasiparticles and collective modes of the order parameter 
encoded in nonlinear optical susceptibilities, as illustrated in Fig.~\ref{fig:2d}(b).


Superconductors host a variety of collective excitations of the order
parameter, including the Higgs mode \cite{Anderson1958, Schmid1968, PekkerVarma2015, ShimanoTsuji2020, Tsuji2024} and, in multiband systems, the Leggett mode associated with relative-phase oscillations
between different condensates \cite{Leggett1966}.
These collective modes typically lie in the terahertz (THz)
frequency range and couple to light nonlinearly in ordinary situations \cite{Tsuji2015, Tsuji2016,Cea2016,Murotani2017,Giorgianni2019,Tsuji2020,linear_response}.
The Higgs mode has been observed through Raman scattering~\cite{Sooryakumar1980,Measson2014,Grasset2018,Grasset2019} and nonlinear
THz spectroscopy~\cite{Matsunaga2013,Matsunaga2014,Matsunaga2017,Katsumi2018},
while the Leggett mode has been identified by Raman spectroscopy in
MgB$_2$~\cite{Blumberg2007}.
THz-2DCS provides a complementary and more systematic means of probing
such collective dynamics by resolving nonlinear excitation pathways
along an additional frequency dimension, thereby helping distinguish
collective-mode resonances from other contributions to the nonlinear
response.

Recent experiments have demonstrated that 2DCS can resolve
nonlinear optical processes that are difficult to disentangle
in conventional one-dimensional spectroscopy \cite{Katsumi2024}.
In particular, the recent application to a multiband superconductor
MgB$_2$~\cite{Katsumi2025} demonstrates the feasibility of probing
the nonlinear optical susceptibility of multiband superconductors,
while also highlighting the challenge of identifying the collective-mode
contributions in the observed 2D spectra.
This raises a fundamental question: 
how do collective-mode resonances
manifest themselves in the 2D spectrum of a multiband superconductor,
and how can they be distinguished from spectral features of other
origins?
Establishing such a correspondence is essential for using 2DCS as a
quantitative probe of superconducting collective dynamics.


On the theoretical side, recent studies have developed frameworks for
interpreting 2DCS spectra in interacting electron systems.
These include 
studies of light-driven states and collective modes in
multiband superconductors~\cite{Mootz2024},
Josephson plasmons in layered superconductors~\cite{GomezSalvador2024}, 
Mott and multi-orbital Hubbard systems~\cite{ChenWerner2025},
antiferromagnetic amplitude modes~\cite{Chen2026}, 
Higgs mode in disordered superconductors~\cite{Tsuji2026},
and a general many-body diagrammatic formulation~\cite{Fiore2026}.

While these studies have established the theoretical treatment of
2DCS in interacting electron systems, how a specific collective-mode (such as Leggett mode)
resonance is encoded in the two-dimensional frequency plane and can be
distinguished from other nonlinear spectral features remains an
open issue.


In this work, we investigate how the Leggett mode appears as a spectral feature in
THz two-dimensional coherent spectroscopy of a two-band superconductor based on a model in Ref.~\cite{Murotani2017}.
Using the Anderson-pseudospin formalism \cite{Anderson1958}, we numerically calculate the nonlinear
dynamics and the resulting 2D spectra. 
In addition to the conventional fundamental- and third-harmonic
features, the calculated spectra exhibit characteristic side peaks
associated with the Leggett-mode frequency.

To elucidate the origin of these features, we derive an analytical
expression for the third-order nonlinear current and identify the
resonance conditions associated with the collective-mode poles of the
nonlinear susceptibility.
We show that pronounced features emerge where these resonance conditions
overlap with regions of spectral weight determined by the incident THz
pulses, providing a systematic interpretation of the otherwise complex
2D spectra.
We further show that these sideband signatures are robust against quasiparticle dissipation by incorporating a phenomenological Gilbert-type relaxation~\cite{damping} of the pseudospins toward equilibrium, which damps the persistent oscillations characteristic of the collisionless BCS limit while leaving the Leggett sidebands intact at their correct spectral positions.
To suggest an experimental route to verify the prediction we show that, by varying the central frequency of the driving pulses, the collective-mode resonances can be distinguished
from pulse-induced spectral structures.
Our results establish THz-2DCS as a powerful approach for identifying
collective-mode dynamics in multiband superconductors and for accessing
information beyond conventional third-harmonic generation and
pump--probe spectroscopy.



The paper is organized as follows. In Sec.~II, we introduce the two-band BCS model and the Anderson-pseudospin formalism, characterize the Leggett mode and its temperature dependence, and present the numerically simulated 2D coherent spectra. In Sec.~III, we develop an analytical interpretation of the 2D map based on the third-order nonlinear susceptibility, deriving the collective-mode resonance conditions and comparing them with the numerics. 
In Sec.~IV, we assess the robustness of the Leggett sideband features against dissipation by including a Gilbert-type pseudospin relaxation.
Finally, in Sec.~V we summarize our results and comment on their experimental relevance for $\text{MgB}_2$, for which THz-2DCS measurements in this regime have already been reported in the experiment~\cite{Katsumi2025}.

\section{2D coherent spectroscopy in a two-band superconductor}

In this section, we formulate our theoretical treatment of 
the THz two-dimensional coherent spectroscopy
for a two-band superconductor.
We first introduce a minimal two-band BCS model and its Anderson
pseudospin representation, using parameters representative of the
characteristic energy scales of MgB$_2$.
We then characterize its collective modes with particular emphasis on
the Leggett mode, and present numerical 2D spectra obtained from
the nonlinear dynamics driven by two THz pulses.

\subsection{Model}\label{sec:model}

Motivated by the experiment for MgB$_2$, we adopt a minimal two-band model of BCS superconductors in which the
two bands are loosely associated with the $\sigma$ and $\pi$ bands.
The model parameters are chosen to be broadly representative of
MgB$_2$, 
while keeping the model tractable for time-domain simulations.
The starting point is the mean-field Hamiltonian,
\begin{equation}
\mathcal{H}_0=\sum_{i \boldsymbol{k} \sigma} \xi_{i \boldsymbol{k}} 
c_{i \boldsymbol{k} \sigma}^{\dagger} c_{i \boldsymbol{k} \sigma}
+\sum_{i \boldsymbol{k}} \psi_i c_{i-\boldsymbol{k} \uparrow}^{\dagger} 
c_{i \boldsymbol{k} \downarrow}^{\dagger}+\text{h.c.},
\label{eq:mgb2_H}
\end{equation}
where $\xi_{i \boldsymbol{k}}=s_i\!\left(\boldsymbol{k}^2/2m_i 
-\epsilon_{F_i}\right)$ is the parabolic dispersion of band $i=1,2$, 
$\epsilon_{F_i}$ is the Fermi energy, and $m_i$ is the effective mass. The 
sign factor $s_i=\pm 1$ distinguishes electron-like bands from hole-like 
ones, and $\psi_i$ is the superconducting order parameter of band 
$i$.

To describe the dynamics driven by an external THz field, it is 
convenient to recast this Hamiltonian in the pseudospin language \cite{Anderson1958}. 
For each Cooper-pair degree of freedom $(\boldsymbol{k},i)$, we define
the Anderson pseudospin as
\begin{align}
\boldsymbol{s}_{\boldsymbol{k}}^{i}
=
\frac{1}{2}
\Psi_{i\boldsymbol{k}}^\dagger
\boldsymbol{\sigma}
\Psi_{i\boldsymbol{k}},
\end{align}
where
\begin{align}
\Psi_{i\boldsymbol{k}}
=
\begin{pmatrix}
c_{i\boldsymbol{k}\uparrow}\\
c_{i-\boldsymbol{k}\downarrow}^{\dagger}
\end{pmatrix}
\end{align}
is the Nambu spinor, and $\boldsymbol{\sigma}=(\sigma_1, \sigma_2, \sigma_3)$ are the Pauli matrices.
Using the pseudospin notation, the Hamiltonian is written as
\begin{align}
H = \sum_{i \boldsymbol{k}} 
\boldsymbol{b}_{\boldsymbol{k}}^{i} \cdot 
\boldsymbol{s}_{\boldsymbol{k}}^{i},
\end{align}
where
\begin{align}
\boldsymbol{b}_{\boldsymbol{k}}^{i} = 
\bigl(-2\psi_i^{\prime},\,2\psi_i^{\prime\prime},\,
2\xi_{i\boldsymbol{k}} + e^2 A^2(t)/2m\bigr)
\label{eq:psudo}
\end{align}
is an effective pseudo-magnetic field around which the pseudospins are precessing,
$A(t)$ is the vector potential of the driving field, and 
$\psi_i^\prime$ and $\psi_i^{\prime\prime}$ are the real and imaginary parts of $\psi_i$.
The order parameters are determined self-consistently by the coupled 
gap equations,
\begin{equation}
\psi_i = \sum_{j \boldsymbol{k}} U_{ij}
\bigl\langle s_{1\boldsymbol{k}}^{j} - 
i s_{2\boldsymbol{k}}^{j}\bigr\rangle,
\end{equation}
where the pairing interaction couples the two bands through the 
matrix
\begin{equation}
U_{ij} = 
\begin{pmatrix}
U_{11} & \nu\, U_{11} \\
\nu\, U_{11} & U_{22}
\end{pmatrix}.
\end{equation}
The off-diagonal entries are controlled by the dimensionless 
interband coupling parameter $\nu$, which plays a central role in 
determining the collective mode spectrum.

\subsection{Collective modes}

Since our bands are parabolic and our system is in the clean limit, both the Higgs amplitude modes do not couple to the electromagnetic field: their optical matrix elements vanish identically in the continuum limit. By choosing $\nu \neq 0$, the relative phase between the two condensates becomes a dynamical degree of freedom, the Leggett mode. 

The Leggett mode frequency can be determined analytically within the 
random phase approximation (RPA)~\cite{Leggett1966,Murotani2017,Giorgianni2019}, which accurately captures 
the collective mode dispersion in the long-wavelength limit. The mode 
frequency $\omega_L$ is the solution of the self-consistency equation
\begin{equation}
\omega_L^2 = L^2(\omega_L),
\label{eq:leggett_self}
\end{equation}
where
\begin{equation}
L^2(\omega) = \kappa\,
\frac{N_1 F_1(\omega) + N_2 F_2(\omega)}
{N_1 N_2 F_1(\omega) F_2(\omega)},
\end{equation}
with $\kappa \equiv 8\Delta_1\Delta_2 U_{12}/\det\hat{U}$ ($U_{12}=\nu U_{11}$)
and $N_i$ the density of states at the Fermi level of band $i$. The functions
\begin{align}
F_i(\omega) = 4\Delta_i^2 
\int_{-\omega_C}^{\omega_C} d\xi\;
\frac{\tanh\bigl(\sqrt{\xi^2+\Delta_i^2}/2T\bigr)}
{\sqrt{\xi^2+\Delta_i^2}
\bigl[4\Delta_i^2 + 4\xi^2 - (\omega+i\delta)^2\bigr]}
\end{align}
are the pair-fluctuation response functions of the individual bands, 
each exhibiting a threshold at $2\Delta_i$ set by the onset of 
quasiparticle pair-breaking.
Here $\omega_C$ is the cut-off frequency, $T$ is the temperature, and $\delta$ is a positive infinitesimal constant.

\begin{figure}[t!] 
    \centering 
 \includegraphics[scale=0.5]{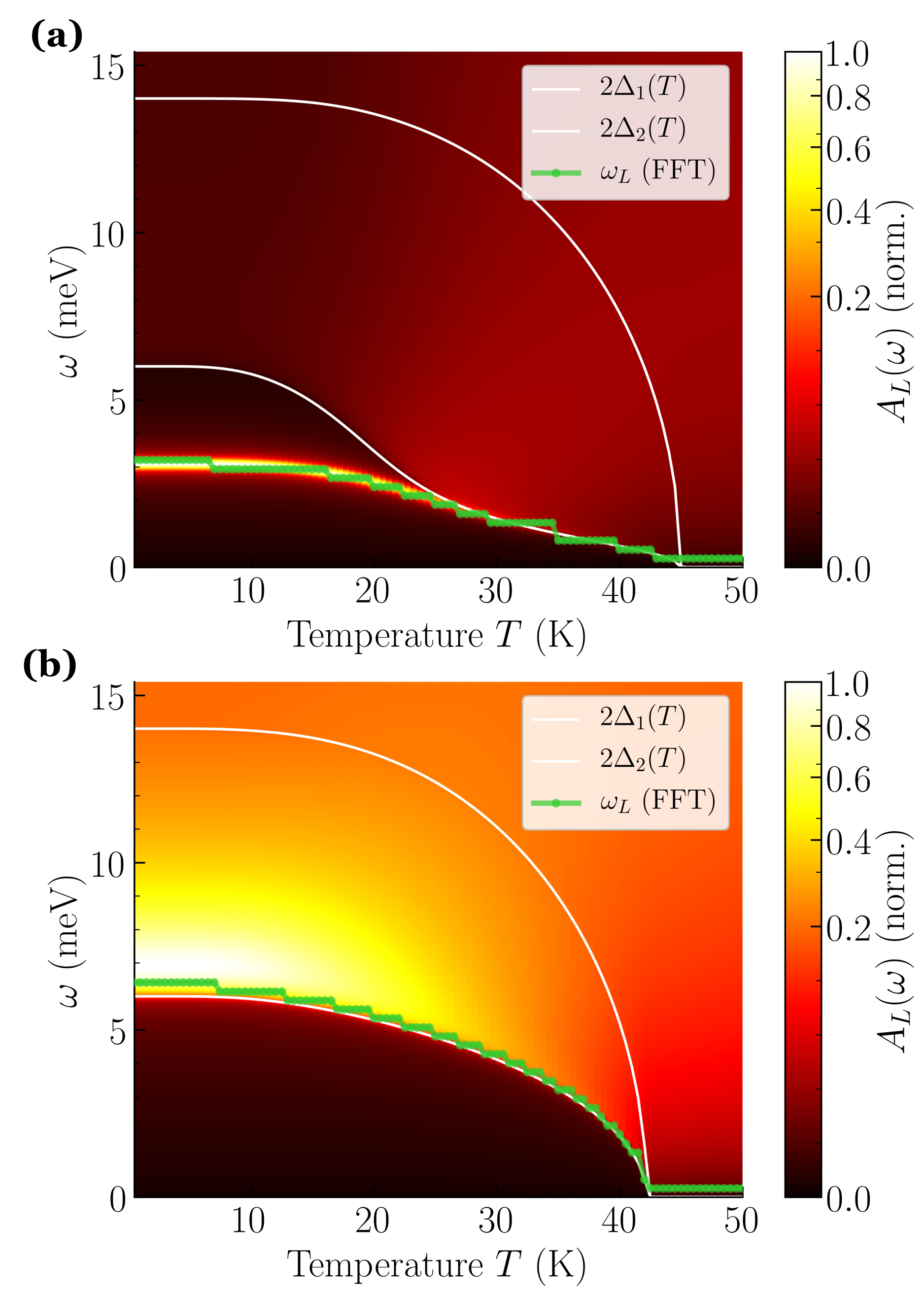} 
    \caption{
    Temperature dependence of the superconducting gaps $2\Delta_1$ and
    $2\Delta_2$ and the Leggett-mode frequency $\omega_L$ for the model in Eq.~\eqref{eq:mgb2_H} with the interband coupling parameter (a) $\nu=0.02$ (in-gap Leggett mode) and (b) $\nu=0.32$ (between-gap Leggett mode).
    The Leggett-mode frequency extracted from the pseudospin dynamics (green lines) is
    overlaid on the spectral function (color map) obtained from Eq.~\eqref{eq:AL}.
    The other parameters are
    $\Delta_1=3$~meV, $\Delta_2=7$~meV,
    $\epsilon_{F,1}=1.6$~eV, $\epsilon_{F,2}=0.7$~eV,
    $m_1=0.85m_e$, $m_2=1.38m_e$, $\epsilon_D=50$~meV,
    $s_1=1$, and $s_2=-1$, chosen to be representative of MgB$_2$.
    }
    \label{fig:leggett}
\end{figure}

In the physically transparent limit of the weak interband coupling, 
Eq.~\eqref{eq:leggett_self} admits the closed-form solution
\begin{equation}
\omega_L^2 \simeq 
\frac{4\Delta_1\Delta_2\,U_{12}}{\det\hat{U}}\,
\frac{N_1 + N_2}{N_1 N_2},
\label{eq:leggett}
\end{equation}
placing the mode below the smaller of the two gaps. In this 
in-gap regime, the Leggett mode is long-lived and gives rise to a sharp 
resonance.
To track $\omega_L$ as a function of temperature, we define the 
Leggett-mode spectral function
\begin{equation}
A_L(\omega) = -\frac{1}{\pi}\,\mathrm{Im}
\left[\frac{\kappa^{-1}L^2(\omega)}{L^2(\omega) - \omega^2}\right],
\label{eq:AL}
\end{equation}
whose peak position identifies $\omega_L(T)$.
Figure~\ref{fig:leggett}(a) displays $A_L(\omega)$ as a color map for
the coupling parameter $\nu=0.02$.
In this regime, the Leggett mode lies below both of the superconducting gap
edges and therefore appears as a sharp, well-defined peak in the
spectrum, well separated from the quasiparticle continuum.

In a more experimentally relevant regime for MgB$_2$, the interband
coupling is sufficiently strong such that the Leggett mode lies between the
two superconducting gap edges,
$2\Delta_\pi < \omega_L < 2\Delta_\sigma$.
%
%

In this regime, the mode overlaps with the quasiparticle continuum
associated with the smaller gap and is therefore intrinsically damped
by pair-breaking processes.
Figure~\ref{fig:leggett}(b) shows the corresponding spectral function
$A_L(\omega)$ for $\nu=0.32$.
As the temperature increases, both superconducting gaps soften and the
Leggett-mode frequency $\omega_L$ decreases accordingly.
At the same time, the spectral overlap between the Leggett mode and the
quasiparticle continuum becomes progressively stronger, leading to a
rapid broadening and suppression of the resonance.
As can be seen in Fig.~\ref{fig:leggett}(b), the Leggett-mode peak becomes
strongly broadened and is effectively washed out above
$T \approx 20\,\mathrm{K}$, well below the bulk 
transition temperature $T_c$.


The Leggett-mode frequency obtained from the RPA spectral function is
compared with that extracted directly from the pseudospin dynamics, as
shown by the green lines in Figs.~\ref{fig:leggett}(a) and (b).
To determine the Leggett-mode frequency from the pseudospin dynamics,
we excite the system with a vector potential of the form
the form
\begin{equation}
    A(t) = A_1\,e^{-t^2/2\tau_1^2}\cos(\Omega_1 t),
    \label{eq:A_leggett_probe}
\end{equation}
where we use a narrow-band pulse with
$\tau_1=1.32$~ps, $\Omega_1=2.18$~THz, and
$A_1=1.7\times10^{-8}~{\rm V\,s\,m^{-1}}$.
The pulse coherently excites the superconducting condensates near the
relevant gap energy scales.
We then solve the coupled Bloch equations for the pseudospins in the two bands.
$\Delta_i(t)$.

From the resulting time evolution of the self-consistent order parameters $\Delta_i(t)$, 
we extract the relative condensate phase
\begin{equation}
    \varphi(t) = \arg(\Delta_1(t) )- \arg(\Delta_2(t)),
\end{equation}
which directly encodes the Leggett mode: 
While the total phase of the condensate is removed by gauge invariance, the relative phase is a 
genuine dynamical degree of freedom that oscillates freely at 
$\omega_L$ once the pulse has passed. Discarding the transient during 
the pulse and Fourier transforming $\varphi(t)$ over the subsequent 
free-oscillation window, the dominant peak in $|\tilde{\varphi}(\omega)|$ 
yields the numerically determined Leggett frequency $\omega_L^{\rm num}$, 
which we compare against the peak of the RPA spectral function 
$A_L(\omega)$ [Eq.~\eqref{eq:AL}].
One can see in Fig.~\ref{fig:leggett} that
the peak position of $A_L(\omega)$ (color map) agrees well with the Leggett-mode frequency extracted
from the pseudospin dynamics (green lines), confirming the consistency
of the two approaches.

\subsection{Numerical results}

To simulate the THz-2DCS response, we drive the system with two
time-delayed pulses of identical shape and carrier frequency.
The total vector potential entering the pseudo-magnetic field is
given by
\begin{equation}
\begin{aligned}
    A(t) &= A_1(t)+A_2(t+\Delta t) \\
    &= A_0\, e^{-\frac{t^2}{2\tau^2}} \cos(\Omega t) \\
    &\quad + A_0\, e^{-\frac{(t-\Delta t)^2}{2\tau^2}} \cos\!\bigl(\Omega(t-\Delta t)\bigr),
\end{aligned}
\label{eq: A(t)}
\end{equation}
where $\Delta t$ denotes the time delay between the two pulses,
$\tau$ their temporal width, 
and $\Omega$ the central frequency.
For each value of $\Delta t$, we numerically solve the coupled Bloch
equations for the pseudospins in the two bands, together with the
self-consistency equations for the time-dependent superconducting
order parameters.
The resulting current as a function of the detection time and pulse
delay forms the time-domain signal from which the 2D spectrum is
constructed.

\begin{figure}[t] 
    \centering 
 \includegraphics[scale=0.43]{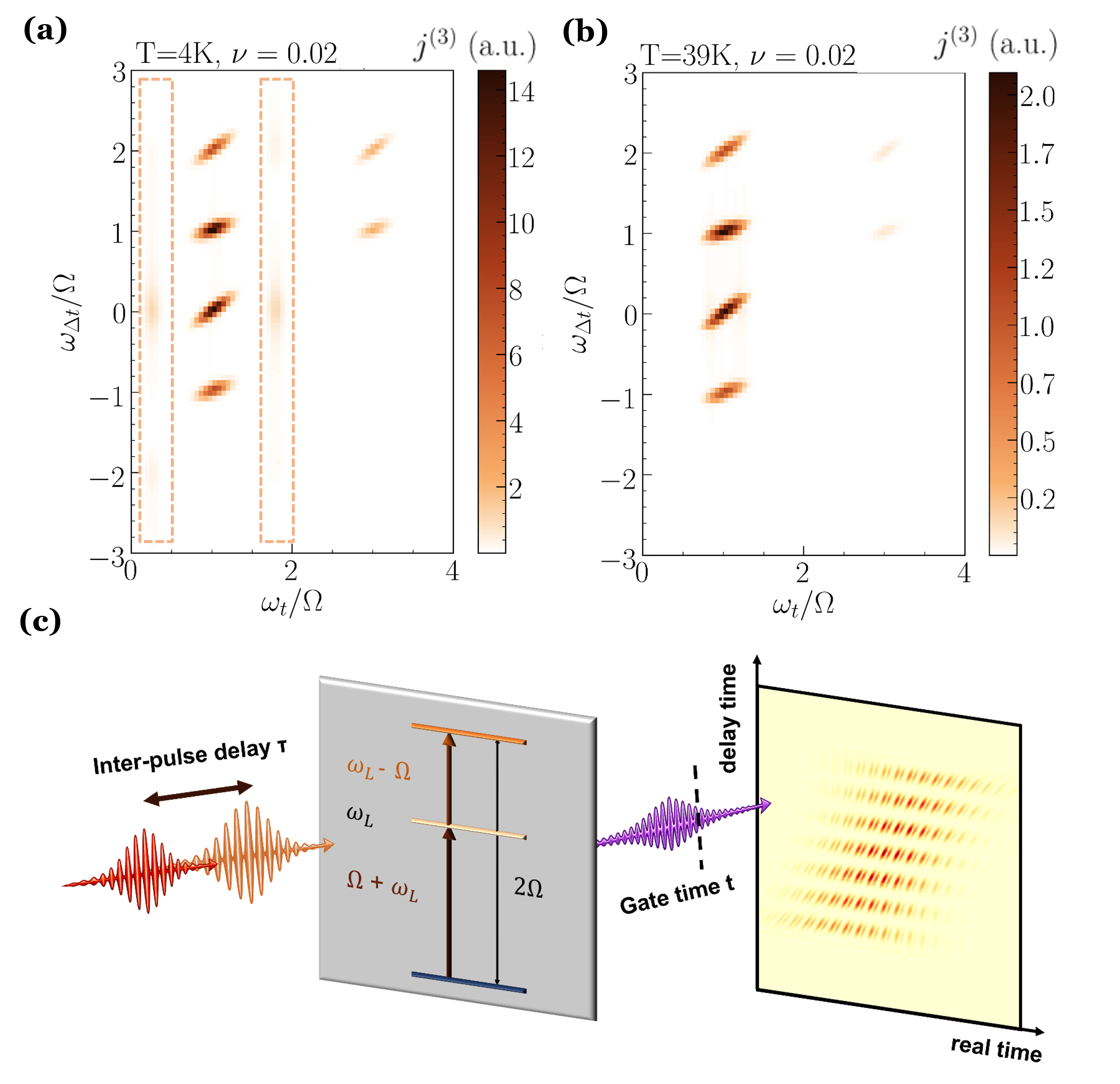} 
    \caption{
    Third-order nonlinear current $j^{(3)}$ of a two-band superconductor calculated within the pseudospin model. 
    (a,b) Two-dimensional spectra in the $(\omega_t,\omega_{\Delta t})$ frequency plane obtained from the model Hamiltonian in Eq.~\eqref{eq:mgb2_H} for $\nu=0.02$, corresponding to an in-gap Leggett mode, at (a) $T=4$~K and (b) $T=39$~K. In (a), characteristic sidebands associated with the Leggett-mode resonance appear at $\omega_t=\pm\omega_L\mp\Omega$, as highlighted by the orange boxes. These sidebands are strongly suppressed upon approaching the critical temperature $T_c=44$~K, as shown in (b).
    (c) Schematic illustration of the nonlinear process responsible for the sidebands. Nonlinear excitation of a collective mode at $\omega_L$ produces spectral features at $\omega_t=\pm\omega_L\mp\Omega$. Consequently, varying the central pulse frequency $\Omega$ shifts the sidebands relative to the intrinsic Leggett-mode frequency, providing a means of identifying their collective-mode origin. The pulse parameters in Eq.~\eqref{eq:A1w} are $A_0=1.7\times10^{-8}~\mathrm{V\,s\,m^{-1}}$, $\Omega=4$~meV, and $\tau=2.64$~ps. The simulations are performed over $t \in[-10,10]$~ps and $\Delta t\in[-8,8]$~ps.
}
    \label{fig:linear}
\end{figure}

The numerically calculated 2D spectra are shown in
Fig.~\ref{fig:linear}.
Comparison with the schematic spectrum in Fig.~\ref{fig:2d}(b) shows
that the numerical results reproduce the characteristic features
expected from third-order frequency mixing, together with additional
structures associated with the collective-mode dynamics.
Figure~\ref{fig:linear}(a) shows the spectrum at $T=4$~K for the
in-gap Leggett-mode regime ($\nu=0.02$).
In addition to the conventional fundamental and third-harmonic features
at $\omega_t=\Omega$ and $3\Omega$, respectively, pronounced sidebands
appear at frequencies shifted from the driving frequency by the
Leggett-mode frequency, $\omega_t=\pm\Omega\pm\omega_L$, as highlighted
by the red dashed boxes.
The frequency $\omega_L$ extracted from these sidebands agrees with the
Leggett-mode frequency independently determined from the relative-phase
dynamics discussed above.

The physical picture underlying these sidebands is illustrated
schematically in Fig.~\ref{fig:linear}(c).
The process can be understood in two stages.
First, the pump pulse nonlinearly drives the two superconducting
condensates.
Since the vector potential enters the pseudo-magnetic field through
$A^2(t)$, the electromagnetic field perturbs both the amplitude and
phase sectors of the order parameters.
In the presence of the interband coupling $U_{12}$, the relative phase
$\varphi_1-\varphi_2$ constitutes a collective dynamical degree of
freedom, and the pump can induce a coherent Leggett oscillation at
$\omega_L$.
After the pump has passed, this relative-phase coherence persists as a
free oscillation, as directly verified from the late-time pseudospin
dynamics and its Fourier spectrum.

In the second stage, the probe pulse arrives after a delay $\Delta t$
and interacts with the condensate while the Leggett oscillation is
present.
The resulting time-dependent superconducting state modulates the
nonlinear optical response at $\omega_L$, allowing the probe frequency
$\Omega$ to mix with the collective-mode frequency.
This frequency mixing produces sidebands at
$\omega_t=\Omega\pm\omega_L$, together with their symmetry-related
counterparts at negative frequencies, as illustrated in
Fig.~\ref{fig:linear}(c).
The process is reminiscent of Stokes and anti-Stokes frequency mixing
with a coherent Raman-active excitation and is closely related to the
sum-frequency mechanism discussed by Giorgianni
\textit{et al.}~\cite{Giorgianni2019}.

The additional frequency dimension of 2DCS provides further information
on this process.
A double Fourier transform with respect to the detection time $t$ and
the inter-pulse delay $\Delta t$ resolves the signal in the
$(\omega_t,\omega_{\Delta t})$ plane, allowing the Leggett-related
sidebands to be separated from the conventional harmonic features.
The resulting sideband structure therefore provides a characteristic
two-dimensional signature of the Leggett coherence that is not available
from the detection-frequency dependence alone.

The temperature dependence in Fig.~\ref{fig:linear}(b) further supports
this interpretation.
With increasing temperature, the superconducting gaps decrease and the
Leggett-mode frequency softens accordingly.
At the same time, the Leggett-related spectral weight is progressively
suppressed, causing the sideband features to weaken and eventually
disappear as the superconducting transition is approached.
This behavior contrasts with the more persistent harmonic peaks ($\omega_t=\Omega, 3\Omega$)
and provides an additional indication that the sidebands originate from
a collective excitation of the superconducting condensate.

\section{Resonance structure of the 2D coherent spectroscopy map}

To interpret the features observed in the numerical 2D map, we consider 
the two-pulse excitation scheme described in Sec.~\ref{sec:model}, where 
both pulses share the same carrier frequency $\Omega$ and are separated 
by a delay time $\tau$. 
Here we make one simplification to obtain an analytical expression for the 2D spectroscopy signal: instead of the Gaussian envelope of Eq.~(\ref{eq: A(t)}), we use a step-function envelope. Each pulse then has a single well-defined onset, the inter-pulse delay is here denoted by $\tau$ rather than $\Delta t$.
Then the total vector potential is given by
\begin{equation}
  A(t) = A_{1}\cos(\Omega t)\,\theta(t)
       + A_{2}\cos\!\bigl(\Omega(t+\Delta t)\bigr)\,\theta(t+\Delta t),
  \label{eq:Avt}
\end{equation}
where $\theta(t)$ is the Heaviside step function encoding the finite 
switch-on time of each pulse. 

The Fourier transform of the first pulse then reads
\begin{equation}
\begin{aligned}
  A_{1}(\omega) =\;& \frac{iA_{1}}{2}
    \left[\mathcal{P}\!\left(\frac{1}{\omega+\Omega}\right)
         +\mathcal{P}\!\left(\frac{1}{\omega-\Omega}\right)\right]\\
  +\;& \frac{\pi A_{1}}{2}
    \left[\delta(\omega+\Omega)+\delta(\omega-\Omega)\right],
\end{aligned}
  \label{eq:A1w}
\end{equation}
where $\mathcal{P}$ denotes the Cauchy principal value. The key 
observation is that, because the pulse is switched on at a finite time, 
the spectrum $A_1(\omega)$ is not confined to the frequency $\pm\Omega$ alone: alongside the 
$\delta$-function spikes at the driving frequency, the step-function 
envelope generates principal-value tails that carry spectral weight at 
\emph{all} frequencies. This broad spectral content is physically 
essential: 
to nonlinearly excite the Leggett mode at $\omega_L \neq 
\Omega$, the driving field must provide frequency components whose 
nonlinear combination can reach $\omega_L$. In the idealized 
monochromatic limit, the spectrum collapses to a pure $\delta$-function 
at $\pm\Omega$ and no such frequency combination 
is possible. The finite pulse 
naturally resolves this problem in the same spirit as the sum-frequency 
excitation strategy of Giorgianni \textit{et al.}~\cite{Giorgianni2019}, 
where two pulses were tuned so that $\Omega_1 + \Omega_2 = \omega_L$. 
Here the broad spectrum of the finite pulse provides the necessary 
frequency content without requiring any tuning.

The nonlinear current extracted from the simulation, 
$j_\mathrm{NL} = j_\mathrm{both} - j_\mathrm{pump} - j_\mathrm{probe}$, 
isolates the cross-coupling contributions between the two pulses. Following 
the formalism of Ref.~\cite{Tsuji2026}, we decompose $j_\mathrm{NL}$ into two 
contributions, $j_{\rm NL}=j_{A_1^2 A_2}+j_{A_1 A_2^2}$, 
which are proportional to $A_1^2A_2$ and $A_1A_2^2$, respectively.

For a current $j(t,\Delta t)$ that depends on the probe time $t$ and the delay time $\Delta t$, we define its two-dimensional Fourier transform as
\begin{align}
j(\omega_t, \omega_{\Delta t})
&=
\int_{-\infty}^\infty dt\; e^{i\omega_t t}\int_{-\infty}^\infty d(\Delta t)\, e^{i\omega_{\Delta t} {\Delta t}} j(t, \Delta t).
\end{align}
The two contributions, $j_{A_1^2A_2}(\omega_t, \omega_{\Delta t})$ and $j_{A_1A_2^2}(\omega_t, \omega_{\Delta t})$, 
give rise to distinct sets of resonance conditions in the two-dimensional frequency plane
$(\omega_t, \omega_{\Delta t})$, 
as discussed below.

Using the third-order nonlinear susceptibility, 
\begin{align}
&
\chi^{(3)}(\omega; \omega_1, \omega_2, \omega_3)
\notag
\\
&=
\frac{1}{3!}
\int_{-\infty}^\infty d\bar t d\bar t' d\bar t'' \, 
e^{i\omega t-i(\omega_1\bar t+\omega_2\bar t'+\omega_3\bar t'')}
\notag
\\
&\quad\times
\frac{\delta^3 j(t)}{\delta A(\bar t)\delta A(\bar t')\delta A(\bar t'')}\bigg|_{A=0}
\quad
(\omega=\omega_1+\omega_2+\omega_3),
\label{chi^(3)}
\end{align}
the contribution proportional to $A_1^2A_2$ can be written as
\begin{align}
j_{A_1^2 A_2}\left(\omega_t, \omega_{\Delta t}\right)
&=
3 A_2\left(\omega_{\Delta t}\right) \int_{-\infty}^{\infty} \frac{d \omega}{2 \pi} A_1(\omega) A_1\left(\omega_t-\omega_{\Delta t}-\omega\right)
\notag\\
&\quad \times \chi^{(3)}\left(\omega_t ; \omega, \omega_t-\omega_{\Delta t}-\omega, \omega_{\Delta t}\right).
\label{eq: j_112}
\end{align}
Equation~\eqref{eq: j_112} provides a general expression for the $A_1^2A_2$ contribution to the 2DCS signal. Importantly, $A_2(\omega_{\Delta t})$ appears as an overall prefactor outside the frequency integral. Consequently, the signal is concentrated around the horizontal line $\omega_{\Delta t}\simeq\Omega$, where $\Omega$ denotes the central frequency of the pulse. Similarly, the contribution proportional to $A_1A_2^2$ is given by 
%
\begin{equation}
\begin{aligned}
j_{A_1 A_2^2}\left(\omega_t, \omega_{\Delta t}\right)= & 3 A_1\left(\omega_t-\omega_{\Delta t}\right) \int_{-\infty}^{\infty} \frac{d \omega}{2 \pi} A_2(\omega) A_2\left(\omega_{\Delta t}-\omega\right) \\
& \times \chi^{(3)}\left(\omega_t ; \omega, \omega_{\Delta t}-\omega, \omega_t-\omega_{\Delta t}\right).
\end{aligned}
\label{eq:current_chi2}
\end{equation}
In this case, $A_1(\omega_t-\omega_{\Delta t})$ appears as an overall prefactor, so that the signal is concentrated around 
$\omega_t-\omega_{\Delta t}\simeq\Omega$. 
This produces a feature parallel to the diagonal $\omega_t=\omega_{\Delta t}$ in the two-dimensional frequency plane.

\begin{figure}[t]
\centering
\includegraphics[width=7cm]{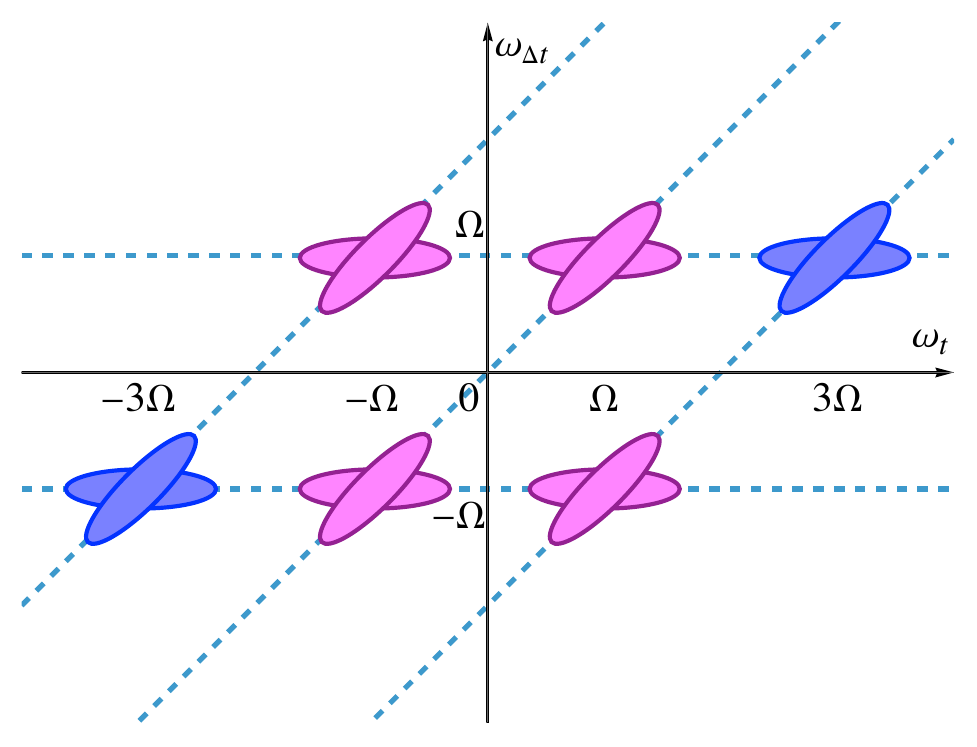}
\caption{
Schematic illustration of the spectral features in 
$j_{A_1^2A_2}(\omega_t,\omega_{\Delta t})$ for narrow-band pulses.
Extended spectral weight appears along the horizontal lines 
$\omega_{\Delta t}=\pm\Omega$ and the diagonal lines 
$\omega_t-\omega_{\Delta t}=0,\pm2\Omega$, as indicated by the dashed lines.
Their intersections produce the characteristic peaks of the 2D spectrum.
The elongated shapes schematically illustrate the finite spectral width 
arising from the principal-value components of the pulse spectrum.}
\label{fig}
\end{figure}

For $j_{A_1^2 A_2}$, substituting Eq.~\eqref{eq:A1w} into Eq.~\eqref{eq: j_112}
and evaluating 
the contributions from both the $\delta$-function and principal-value parts of the 
field, we obtain
\begin{widetext}
\begin{align}
j_{A_1^2A_2}(\omega_t, \omega_{\Delta t})
&=
-\frac{3}{4}
A_1^2A_2(\omega_{\Delta t})
\int_{-\infty}^\infty \frac{d\omega}{2\pi}
\mathcal P
\bigg(\frac{1}{\omega_t-\omega_{\Delta t}+2\Omega}\bigg(\frac{1}{\omega+\Omega}+\frac{1}{\omega_t-\omega_{\Delta t}-\omega+\Omega}\bigg)
\notag
\\
&\quad
+\frac{1}{\omega_t-\omega_{\Delta t}}\bigg(\frac{1}{\omega+\Omega}+\frac{1}{\omega_t-\omega_{\Delta t}-\omega-\Omega}\bigg)\bigg)
\chi^{(3)}(\omega_t; \omega, \omega_t-\omega_{\Delta t}-\omega, \omega_{\Delta t})
\notag
\\
&\quad
-\frac{3}{4}
A_1^2A_2(\omega_{\Delta t})
\int_{-\infty}^\infty \frac{d\omega}{2\pi}
\mathcal P
\bigg(
\frac{1}{\omega_t-\omega_{\Delta t}}\bigg(\frac{1}{\omega-\Omega}+\frac{1}{\omega_t-\omega_{\Delta t}-\omega+\Omega}\bigg)
\notag
\\
&\quad
+\frac{1}{\omega_t-\omega_{\Delta t}-2\Omega}\bigg(\frac{1}{\omega-\Omega}+\frac{1}{\omega_t-\omega_{\Delta t}-\omega-\Omega}\bigg)\bigg)
\chi^{(3)}(\omega_t; \omega, \omega_t-\omega_{\Delta t}-\omega, \omega_{\Delta t})
\notag
\\
&\quad
+\frac{3i}{4}
A_1^2A_2(\omega_{\Delta t})
\bigg(\frac{1}{\omega_t-\omega_{\Delta t}+2\Omega}+\frac{1}{\omega_t-\omega_{\Delta t}}\bigg)
\chi^{(3)}(\omega_t; \omega_t-\omega_{\Delta t}+\Omega, -\Omega, \omega_{\Delta t})
\notag
\\
&\quad
+\frac{3i}{4}
A_1^2A_2(\omega_{\Delta t})
(\frac{1}{\omega_t-\omega_{\Delta t}}+\frac{1}{\omega_t-\omega_{\Delta t}-2\Omega})
\chi^{(3)}(\omega_t; \omega_t-\omega_{\Delta t}-\Omega, \Omega, \omega_{\Delta t})
\notag
\\
&\quad
+\frac{3\pi}{8}A_1^2A_2(\omega_{\Delta t})
(\delta(\omega_t-\omega_{\Delta t})+\delta(\omega_t-\omega_{\Delta t}-2\Omega))
\chi^{(3)}(\omega_t; \Omega, \omega_t-\omega_{\Delta t}-\Omega, \omega_{\Delta t})
\notag
\\
&\quad
+\frac{3\pi}{8}A_1^2A_2(\omega_{\Delta t})
(\delta(\omega_t-\omega_{\Delta t}+2\Omega)+\delta(\omega_t-\omega_{\Delta t}))
\chi^{(3)}(\omega_t; -\Omega, \omega_t-\omega_{\Delta t}+\Omega, \omega_{\Delta t}).
\label{j^(3)}
\end{align}
Although the resulting expression is rather involved, its structure can be understood in a simple way. The frequency dependence is governed by combinations of factors of the form
\begin{align}
&
\bigg(\frac{1}{\omega_{\Delta t}+\Omega} \mbox{ or } \frac{1}{\omega_{\Delta t}-\Omega} \mbox{ or } \delta(\omega_{\Delta t}+\Omega)
\mbox{ or } \delta(\omega_{\Delta t}-\Omega)\bigg)
\notag
\\
&\times
\bigg(\frac{1}{\omega_t-\omega_{\Delta t}} \mbox{ or } \frac{1}{\omega_t-\omega_{\Delta t}+2\Omega} \mbox{ or } \frac{1}{\omega_t-\omega_{\Delta t}-2\Omega}
 \mbox{ or } \delta(\omega_t-\omega_{\Delta t})
\mbox{ or } \delta(\omega_t-\omega_{\Delta t}+2\Omega)
\mbox{ or } \delta(\omega_t-\omega_{\Delta t}-2\Omega)\bigg).
\end{align}
\end{widetext}
This structure has an important consequence for the two-dimensional spectrum. The products of the $\delta$-function terms generate discrete peaks at the intersections of $\omega_{\Delta t}=\pm\Omega$ and $\omega_t-\omega_{\Delta t}=0, \pm 2\Omega$,
corresponding to the conventional discrete spots 
at $(\omega_{\Delta t}, \omega_t)=(\pm\Omega, \pm\Omega), \pm(3\Omega, \Omega)$ for monochromatic pulses.
In contrast, terms containing one or more principal-value factors are not confined to these intersections. Instead, they give rise to spectral weight extending along the horizontal lines
$\omega_{\Delta t}=\pm\Omega$ and the diagonal lines $\omega_t-\omega_{\Delta t}=0, \pm 2\Omega$, 
as illustrated in Fig.~\ref{fig}.

Consequently, even in the narrow-band-pulse regime, the 2D spectrum is not restricted to isolated $\delta$-function peaks. The principal-value components of the pulse spectrum generate extended spectral features and can therefore produce side peaks away from the discrete monochromatic spots. This observation is particularly important when interpreting additional structures in the 2D spectrum. 
An analogous analysis applies to $j_{A_1A_2^2}(\omega_t, \omega_{\Delta t})$.

\begin{figure}[t]
\centering
\includegraphics[width=6cm]{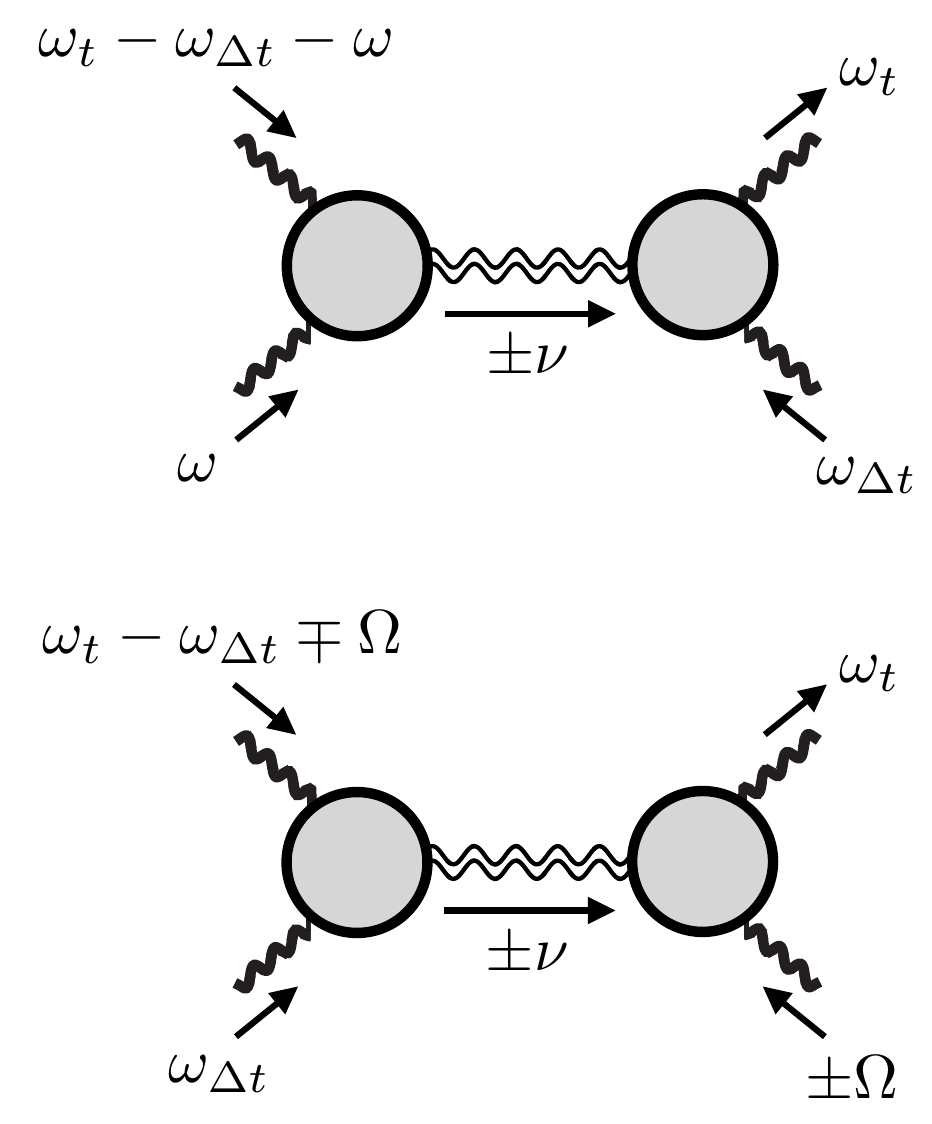}
\caption{
Diagrammatic representation of the collective-mode resonance in the
third-order nonlinear response.
The internal wavy line represents a collective-mode excitation with
frequency $\pm\nu$.
In the upper panel, the three incoming frequencies are
$\omega$, $\omega_t-\omega_{\Delta t}-\omega$, and $\omega_{\Delta t}$,
leading to the resonance condition
$\omega_t-\omega_{\Delta t}=\pm\nu$, which is independent of the integration
frequency $\omega$.
The lower panel illustrates the corresponding process when one of the
incoming frequencies is fixed to $\pm\Omega$ by the $\delta$-function
component of the narrow-band pulse spectrum.}

\label{fig: diagram}
\end{figure}

The nonlinear susceptibility $\chi^{(3)}(\omega; \omega_1, \omega_2, \omega_3)$ 
is expected to exhibit a pole, or more generally a resonance structure,
when the frequency difference between the output and one of the input
frequencies matches the collective-mode frequency $\nu$, i.e.,
$\omega-\omega_i=\pm\nu$ ($i=1,2,3$).
In Eq.~\eqref{j^(3)}, the susceptibility appearing in the frequency
integral takes the form
$\chi^{(3)}(\omega_t; \omega, \omega_t-\omega_{\Delta t}-\omega, \omega_{\Delta t})$. 
The corresponding resonance conditions are therefore
\begin{align}
\omega_t-\omega
&=
\pm\nu,
\\
\omega_{\Delta t}+\omega
&=
\pm\nu,
\\
\omega_t-\omega_{\Delta t}
&=
\pm\nu.
\end{align}
The first two conditions depend explicitly on the integration variable
$\omega$ and are therefore sampled over a range of frequencies in the
convolution with the pulse spectra.
By contrast, the third condition is independent of $\omega$ and directly
defines a resonance line in the two-dimensional frequency plane,
$\omega_t-\omega_{\Delta t}=\pm\nu$.
Thus, a collective mode of frequency $\nu$ can give rise to a
characteristic feature along a line parallel to the diagonal
$\omega_t=\omega_{\Delta t}$.
The corresponding nonlinear process is illustrated in the upper panel
of Fig.~\ref{fig: diagram}.

In addition to the susceptibility appearing in the frequency integral,
Eq.~\eqref{j^(3)} contains terms in which some of the frequency arguments
of $\chi^{(3)}$ are fixed by the $\delta$-function components of the pulse
spectrum.
These terms give rise to three types of collective-mode resonance conditions,
\begin{align}
\omega_t\pm\Omega
&=
\pm\nu,
\\
\omega_{\Delta t}\pm\Omega
&=
\pm\nu,
\\
\omega_t-\omega_{\Delta t}
&=
\pm\nu.
\end{align}
The third condition, $\omega_t-\omega_{\Delta t}=\pm\nu$, coincides with the
resonance condition identified above and therefore gives rise to the same
diagonal feature in the 2D frequency plane.

The first condition instead fixes the frequency $\omega_t$ as
\begin{equation}
\omega_t=\pm\nu\mp\Omega,
\end{equation}
and can therefore generate additional side peaks associated with the
collective-mode resonance.
The corresponding nonlinear process is illustrated in the lower panel
of Fig.~\ref{fig: diagram}.

Although the second condition formally gives another possible resonance,
$\omega_{\Delta t}=\pm\nu\mp\Omega$, its contribution is strongly constrained
by the overall factor $A_2(\omega_{\Delta t})$ in
$j_{A_1^2A_2}(\omega_t,\omega_{\Delta t})$.
For a narrow-band pulse, this factor restricts the appreciable spectral
weight to $\omega_{\Delta t}\simeq\pm\Omega$.
Consequently, the condition
$\omega_{\Delta t}\pm\Omega=\pm\nu$ generally lies outside the region of
substantial pulse spectral weight, except near the special conditions
$\nu\simeq0$ or $\nu\simeq2\Omega$.
We therefore expect the collective-mode signatures associated with
$\omega_t\pm\Omega=\pm\nu$ and
$\omega_t-\omega_{\Delta t}=\pm\nu$ to be considerably more prominent in the
2D spectrum.

\begin{figure}[t]
\centering
\includegraphics[width=8cm]{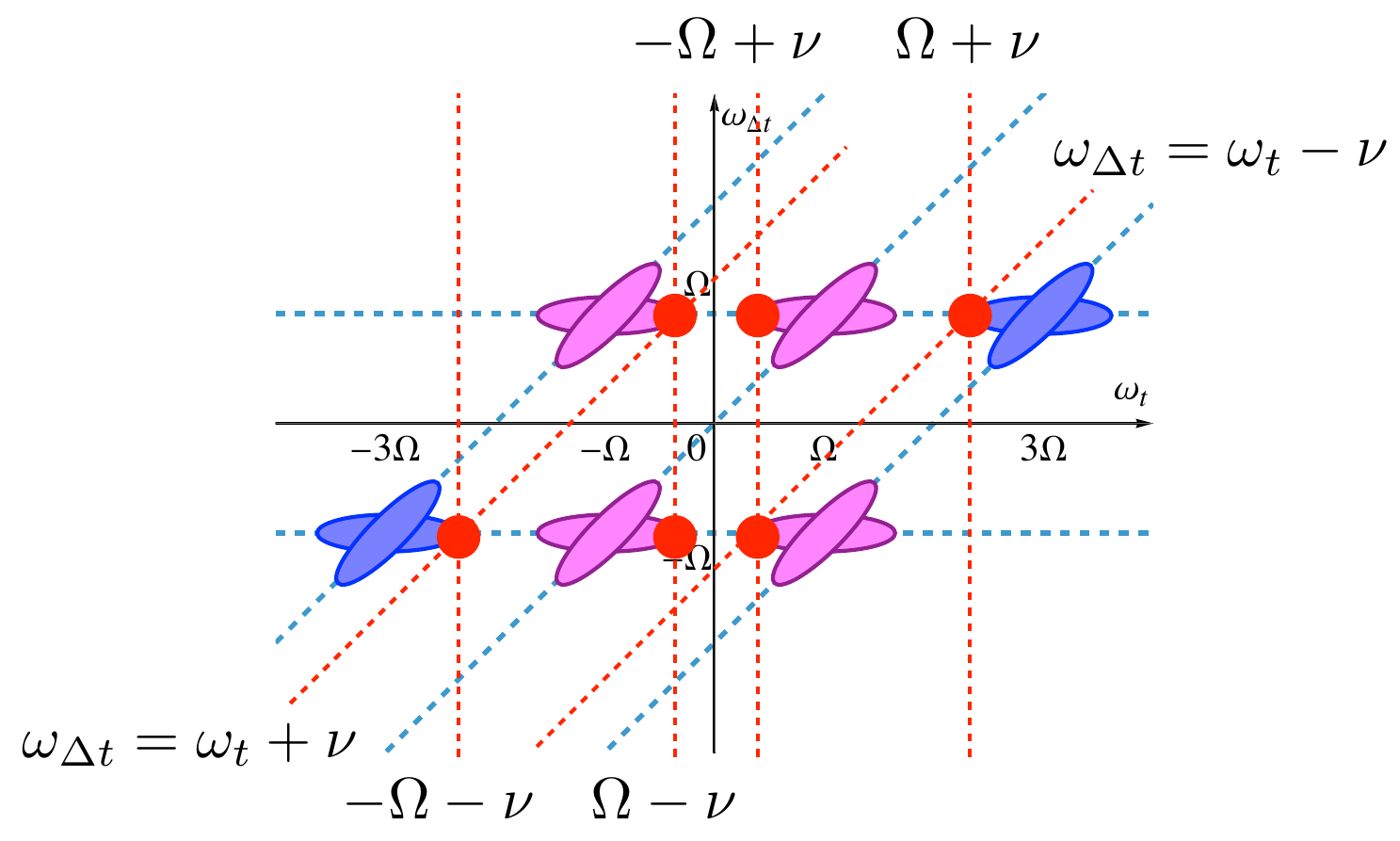}
\caption{
Schematic illustration of the collective-mode resonance conditions in
the two-dimensional frequency plane for the $A_1^2A_2$ contribution.
The blue dashed lines indicate the spectral features of
$j_{A_1^2A_2}(\omega_t,\omega_{\Delta t})$ imposed by the narrow-band pulse
spectrum, while the red dashed lines represent the collective-mode
resonance conditions
$\omega_t-\omega_{\Delta t}=\pm\nu$ and
$\omega_t\pm\Omega=\pm\nu$.
The red circles mark the hot spots where the collective-mode resonance
lines overlap with regions of appreciable 2DCS spectral weight.
These intersections therefore indicate the frequencies at which the
collective-mode response is expected to be resonantly enhanced.}
\label{fig: resonance}
\end{figure}

In Fig.~\ref{fig: resonance}, we summarize the collective-mode resonance
conditions in the two-dimensional frequency plane.
The red dashed lines represent the resonance conditions derived above,
\begin{align}
\omega_t-\omega_{\Delta t} &= \pm\nu,\\
\omega_t\pm\Omega &= \pm\nu.
\end{align}
These conditions alone, however, do not necessarily imply an observable
enhancement of the 2DCS signal.
For a collective-mode resonance to produce a pronounced spectral feature,
the resonance line must overlap with a region where the underlying
$j_{A_1^2A_2}(\omega_t,\omega_{\Delta t})$ signal carries appreciable spectral
weight.

As discussed above, for narrow-band pulses the spectral weight of
$j_{A_1^2A_2}$ is concentrated around the horizontal lines
$\omega_{\Delta t}=\pm\Omega$ and the diagonal lines
$\omega_t-\omega_{\Delta t}=0,\pm2\Omega$, with finite-width extensions arising
from the principal-value components of the pulse spectrum.
The intersections of these pulse-induced spectral features with the
collective-mode resonance lines therefore define particularly favorable
regions for observing the collective-mode response.
These resonantly enhanced regions are indicated by the red markers in
Fig.~\ref{fig: resonance}.

In this sense, the positions of the hot spots are determined by two
independent ingredients: the kinematic structure imposed by the pulse
spectrum and the intrinsic resonance condition of the nonlinear
susceptibility.
The former depends explicitly on the central driving frequency $\Omega$,
whereas the latter contains the collective-mode frequency $\nu$.
Consequently, varying $\Omega$ shifts the pulse-induced structures and
some of the resonance lines relative to one another, providing a useful
means of distinguishing genuine collective-mode resonances from
nonresonant spectral features generated by the finite pulse width.

A similar set of collective-mode resonances appears in the
$j_{A_1A_2^2}$ contribution.
Rather than deriving the resonance conditions independently, they can be
obtained directly from those of $j_{A_1^2A_2}$ by using the symmetry
relation \cite{Tsuji2026},
\begin{align}
j_{A_1^2A_2}(\omega_t,\omega_{\Delta t})
&=
j_{A_1A_2^2}(\omega_t,\omega_t-\omega_{\Delta t}).
\end{align}
Thus, the resonance structure of $j_{A_1A_2^2}$ follows from that of
$j_{A_1^2A_2}$ under the transformation
$\omega_{\Delta t}\rightarrow\omega_t-\omega_{\Delta t}$ in the two-dimensional
frequency plane.

\begin{figure}[t]
\centering
\includegraphics[width=8cm]{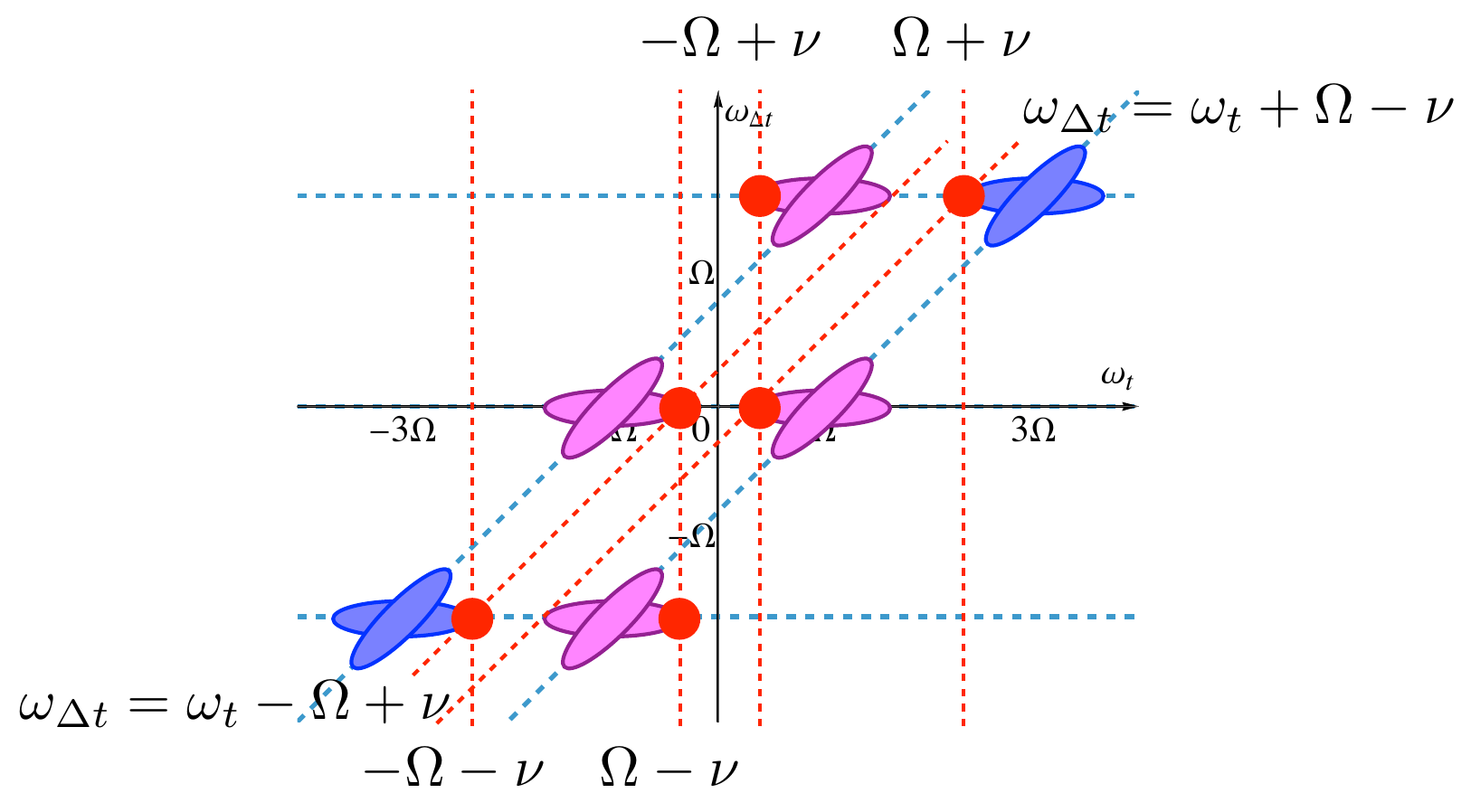}
\caption{Schematic illustration of the collective-mode resonance conditions in
the two-dimensional frequency plane for the $A_1A_2^2$ contribution.
The blue dashed lines indicate the spectral features of
$j_{A_1A_2^2}(\omega_t,\omega_{\Delta t})$ imposed by the narrow-band pulse
spectrum, while the red dashed lines represent the collective-mode
resonance conditions obtained from those of the $A_1^2A_2$ contribution
under the transformation
$\omega_{\Delta t}\rightarrow\omega_t-\omega_{\Delta t}$.
The red circles mark the hot spots where the resonance lines overlap
with regions of appreciable 2DCS spectral weight, indicating the
frequencies at which the collective-mode response is expected to be
resonantly enhanced.}
\label{fig: resonance2}
\end{figure}

\begin{figure}[t]
\centering
\includegraphics[width=6cm]{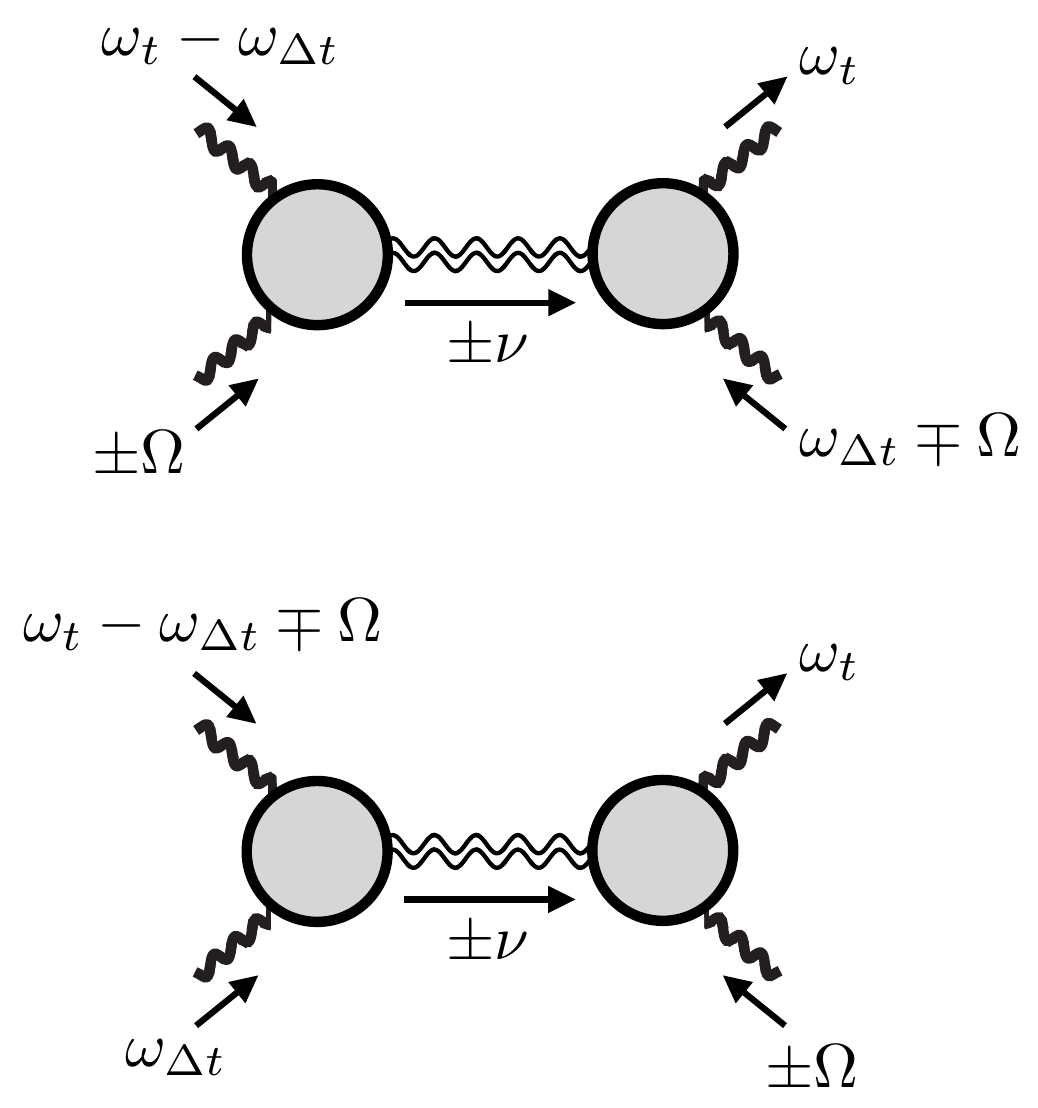}
\caption{
Diagrammatic representation of the collective-mode resonances in the
$A_1A_2^2$ contribution to the third-order nonlinear response.
The internal wavy line represents a collective-mode excitation with
frequency $\pm\nu$.
The upper and lower panels illustrate the nonlinear processes associated
with the resonance conditions obtained from the different frequency
assignments of the two pulses.
The frequencies fixed at $\pm\Omega$ originate from the
$\delta$-function components of the narrow-band pulse spectrum.}
\label{fig: diagram2}
\end{figure}

The resulting resonance conditions are summarized schematically in
Fig.~\ref{fig: resonance2}.
As in the $A_1^2A_2$ contribution, a pronounced collective-mode signal is
expected when a resonance line overlaps with a region in which
$j_{A_1A_2^2}(\omega_t,\omega_{\Delta t})$ carries appreciable spectral weight.
These intersections give rise to a number of characteristic hot spots in
the 2D spectrum, as indicated in Fig.~\ref{fig: resonance2}.
The nonlinear processes responsible for the corresponding collective-mode
resonances are illustrated diagrammatically in
Fig.~\ref{fig: diagram2}.

\begin{figure}[t!] 
    \centering 
 \includegraphics[scale=0.6]{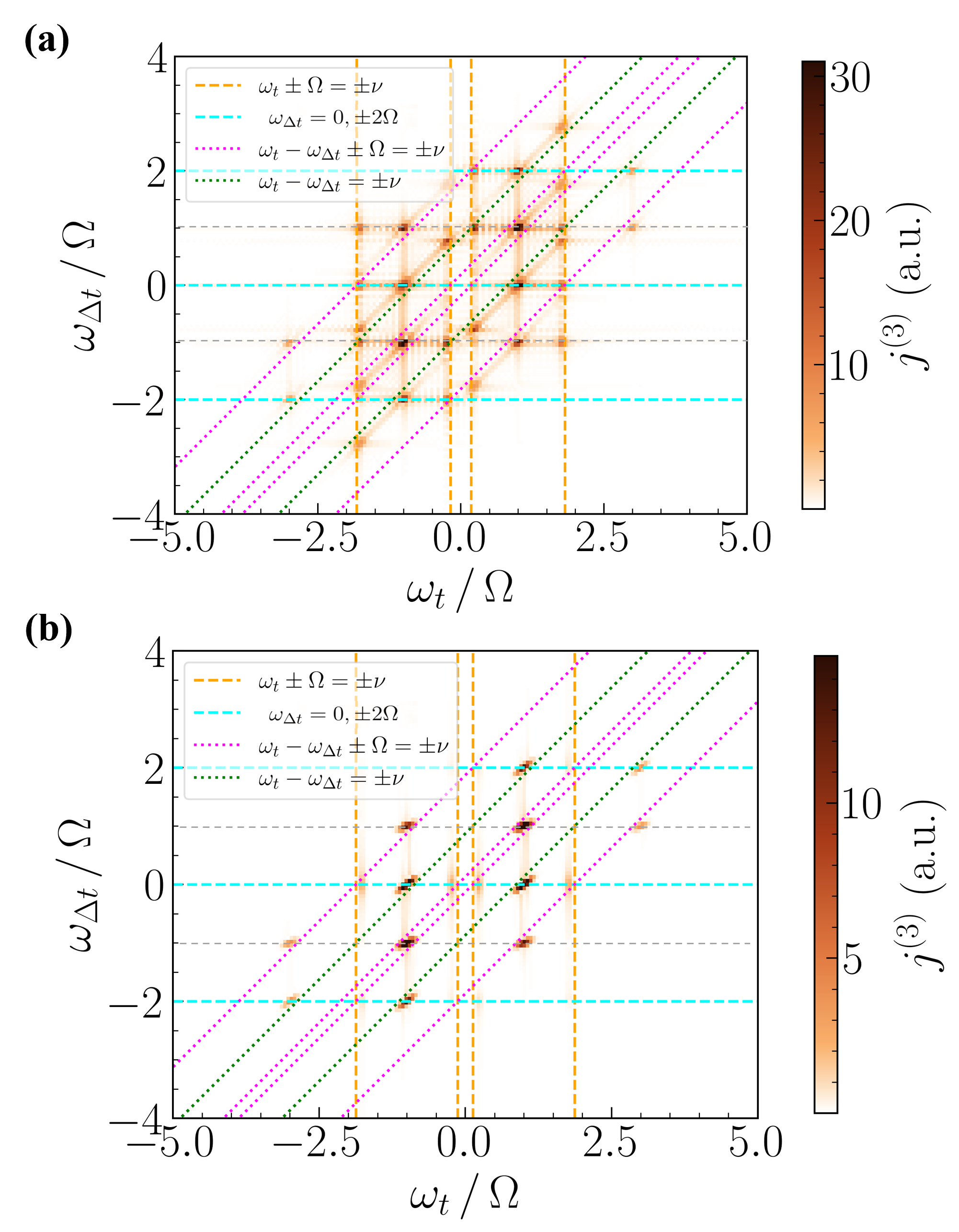} 
\caption{Resonance structure of the 2DCS map: analytics vs.\ numerics.
(a) $|j_{\rm NL}(\omega_t,\omega_{\Delta t})|$ obtained from the
pseudospin equations driven by the step-switched fields of
Eq.~\eqref{eq:Avt}, i.e., the drive used in the analytical treatment.
(b) Same quantity for the Gaussian pulses of Eq.~\eqref{eq: A(t)}.
Both panels are shown on the same linear color scale.
Colored lines: collective-mode resonance conditions,
$\omega_t\pm\Omega=\pm\omega_L$ (orange, dashed),
$\omega_t-\omega_{\Delta t}=\pm\omega_L$ (green, solid; $A_1^2A_2$
contribution), and $\omega_{\Delta t}=\pm\omega_L$ (magenta,
dash-dotted; $A_1A_2^2$ contribution).
Thin grey lines: pulse-induced spectral-weight lines,
$\omega_{\Delta t}=\pm\Omega$ and $\omega_t-\omega_{\Delta t}=0,\pm2\Omega$
($A_1^2A_2$), and $\omega_t-\omega_{\Delta t}=\pm\Omega$ and
$\omega_{\Delta t}=0,\pm2\Omega$ ($A_1A_2^2$).
Enhanced spectral weight appears where the resonance lines cross the
spectral-weight lines.
Parameters: $\Omega=4$~meV, $\omega_L/\Omega\simeq0.8$, interband
coupling $\nu=0.02$, $T=4$~K.}
    \label{fig:theory}
\end{figure}

The underlying amplitude structure is determined primarily by the
frequency content of the two pulses, while the collective-mode poles of
the nonlinear susceptibility introduce additional resonance conditions
in the $(\omega_t,\omega_{\Delta t})$ plane.

An important point is that satisfying a collective-mode resonance
condition alone does not necessarily produce a strong signal.
A pronounced spectral feature is expected only when a resonance line
passes through a region in which the underlying nonlinear response has
appreciable spectral weight.
The intersections between the pulse-induced amplitude structures and
the collective-mode resonance lines therefore define a set of
characteristic hot spots in the 2D frequency plane.
In this way, the locations of the enhanced peaks are determined jointly
by the kinematic constraints imposed by the pulse spectrum and the
intrinsic collective-mode frequency $\nu$.
This prediction is directly confirmed by numerical calculations.
In Fig.~\ref{fig:theory}(a) the numerical 2D spectrum obtained for the pulse shape adopted in the theoretical description, see Eq.\eqref{eq:Avt}, is displayed. We present it as the controlled limit that validates the analytics, not as the experimental scenario. By inspection of the map we see that the hotspots are concentrated near the predicted intersections between the resonance
lines and the regions of appreciable spectral weight.
Furthermore, by comparing  Fig.~\ref{fig:theory}(a) with the numerical 2D spectrum obtained for the finite duration pulses (Eq.\eqref{eq: A(t)}) in Fig.~\ref{fig:theory}(b), remarkably, we notice that, even if a little bit fainted, the features are still where they are supposed to be.

As discussed above, a resonance condition alone does not guarantee a
visible feature: a pronounced peak appears only where a resonance line
crosses a region of appreciable pulse-induced spectral weight. The
positions of these hot spots are therefore fixed by the carrier
frequency $\Omega$ and the collective-mode frequency $\omega_L$, whereas
their intensities are controlled by the spectral content of the pulses.

To test these predictions directly, we solve the pseudospin equations
of Sec.~II with exactly the drive assumed in the analytical treatment,
i.e., the step-switched fields of Eq.~\eqref{eq:Avt}, and extract
$j_{\rm NL}=j_{\rm both}-j_{\rm pump}-j_{\rm probe}$ as before.
An abrupt switch-on of this kind is not realized in experiment; it
serves here as a controlled limit in which the numerics and the
analytics refer to the same drive.
The resulting 2D spectrum is shown in Fig.~\ref{fig:theory}(a), with the
predicted resonance lines overlaid. The dominant features lie at
intersections between the resonance lines and the pulse-induced
spectral-weight lines, for both the $A_1^2A_2$ and the $A_1A_2^2$
contributions. The red circles in Figs.~\ref{fig: resonance} and
\ref{fig: resonance2} mark only a representative subset of these
intersections, chosen for illustration. The numerics show that the rule
itself holds: enhanced spectral weight appears at
essentially every point where a resonance line crosses a region of
appreciable pulse-induced weight, including several intersections not
marked in the schematic figures. Conversely, no pronounced features appear away from such intersections.
 
Figure~\ref{fig:theory}(b) shows the corresponding spectrum for the
Gaussian pulses of Eq.~\eqref{eq: A(t)} with the same carrier frequency.
The hot spots appear at the same positions, but their intensity is
reduced relative to Fig.~\ref{fig:theory}(a).
This is the expected behavior: the positions are set by $\Omega$ and
$\omega_L$ alone, while the intensities depend on how much spectral
weight the drive carries at the frequency combinations that reach
$\omega_L$. The spectrum of an abruptly switched field decays only
algebraically, as $1/(\omega\mp\Omega)$ [Eq.~\eqref{eq:A1w}], whereas
that of a Gaussian envelope decays exponentially on the scale of the
pulse bandwidth. The collective-mode signatures are therefore robust in
position but not in intensity, which directly constrains the choice of
pulse parameters in an experiment.

The agreement demonstrates that the seemingly complicated pattern of
peaks in the 2D spectrum can be understood systematically in terms of
the resonance conditions derived from the third-order nonlinear
susceptibility.
In particular, the additional side peaks away from the conventional
pulse-induced structures can be traced to resonant excitation of the
collective mode rather than to the finite spectral width of the pulses
alone.
Thus, the two-dimensional frequency dependence provides a direct means
of separating genuine collective-mode resonances from spectral features
arising purely from the pulse envelope.

\section{Effects of pseudospin relaxation}\label{sec:relax}

\begin{figure}[t!] 
    \centering 
 \includegraphics[scale=.76]{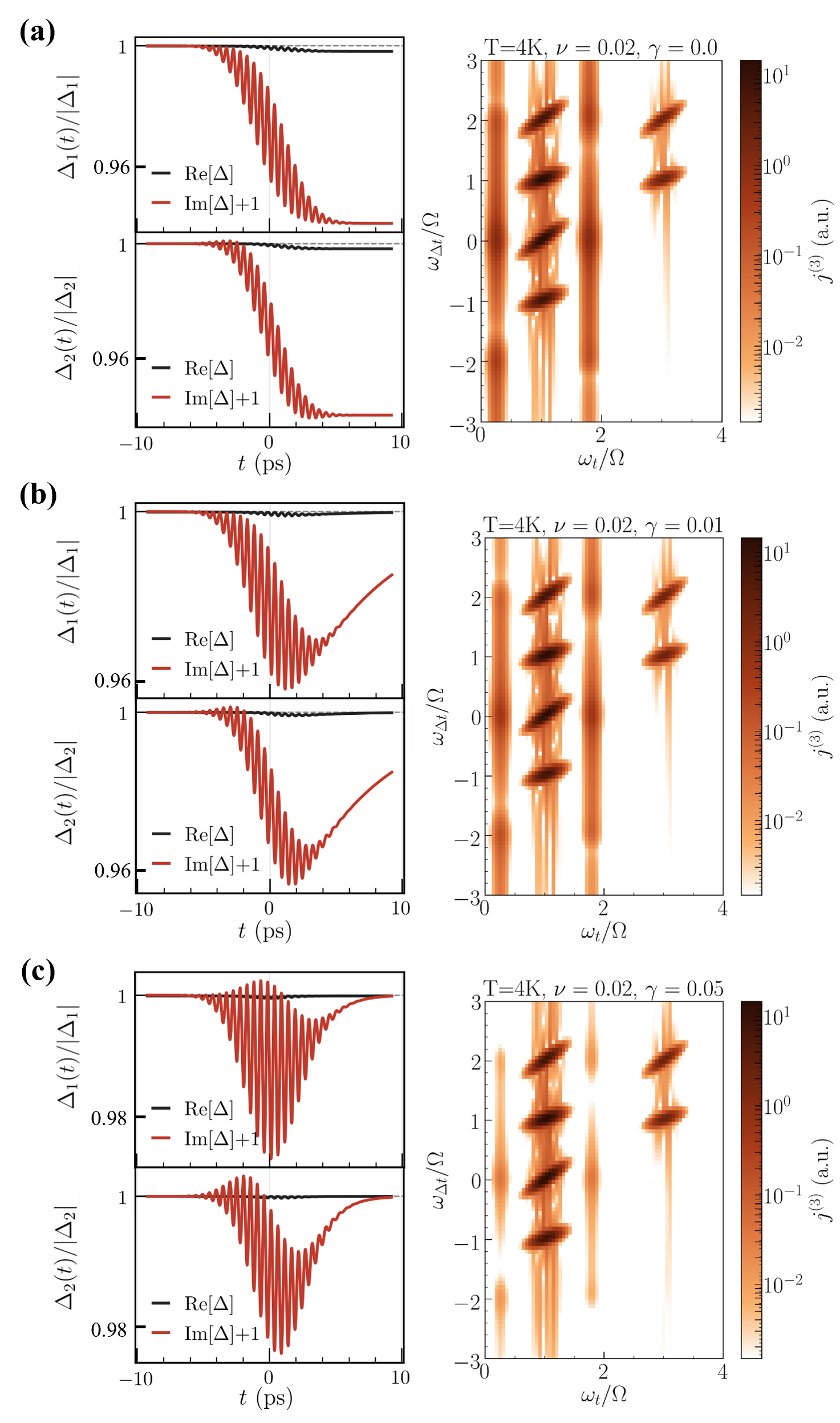} 
    \caption{Pseudospins dynamics with different damping coefficients. Left panels: time evolution of the real and imaginary parts of the two gaps, ${\rm Re}[\Delta_{1,2}(t)]/|\Delta_{1,2}|$ and ${\rm Im}[\Delta_{1,2}(t)]/|\Delta_{1,2}|$, with $|\Delta_{1,2}|$ the equilibrium values of the gaps. 
    Right panels: corresponding 2D spectra shown on a logarithmic scale. (a) Collisionless limit, $\gamma=0$. (b) $\gamma=0.1$~meV. (c) $\gamma=0.5$~meV.}
    \label{fig:relax_gap}
\end{figure}

Dissipation and relaxation are unavoidable in realistic superconductors
and can substantially modify their nonlinear optical response.
For example, impurity scattering is known to play an important role
in the nonlinear THz response of superconductors, strongly affecting
the coupling of electromagnetic fields to collective modes~\cite{Jujo2018,Murotani2019,Silaev2019,Tsuji2020,Seibold2021,Derendorf2024}.

In order to assess the robustness of the predicted sideband features 
against dissipation, we extend the Bloch equations to include a 
phenomenological relaxation of the pseudospins toward their equilibrium 
values. 
We adopt a Gilbert-type damping 
term~\cite{Barankov_2004,Tseng_2026}, 
which modifies the pseudo-magnetic field as
\begin{equation}
    \mathbf{b}_k^{\mathrm{eff}} = \mathbf{b}_k 
    - \gamma\, \hat{\mathbf{b}}_k \times \mathbf{s}_k,
    \label{eq:gilbert}
\end{equation}
so that the equation of motion becomes
\begin{align}
\dot{\mathbf{s}}_k = \mathbf{b}_k \times \mathbf{s}_k 
+ 2\gamma\,(\hat{\mathbf{b}}_k \times \mathbf{s}_k)\times \mathbf{s}_k.
\end{align}
Since $(\hat{\mathbf{b}}_k \times \mathbf{s}_k)\times\mathbf{s}_k$ is 
perpendicular to $\mathbf{s}_k$, the magnitude $|\mathbf{s}_k|$ is 
exactly conserved: the Gilbert term rotates each pseudospin toward the 
local equilibrium direction $\hat{\mathbf{b}}_k$ without changing its 
length. 
Physically, the $z$-component of $\mathbf{s}_k$ 
encodes the quasiparticle occupation,
while the transverse components 
carry the superconducting pair correlation associated with the off-diagonal order 
parameter $\Delta_i$. The Gilbert term drives the 
quasiparticle occupation toward equilibrium and damps the oscillations
of the pair correlations with a single rate $\gamma$. 
This can be viewed as a phenomenological description of dissipative dynamics
with a single parameter of the relaxation rate~\cite{Tseng_2026}.

The collisionless limit ($\gamma=0$) shown in Fig.~\ref{fig:relax_gap}(a)
corresponds to the regime of pure BCS mean-field dynamics in the absence 
of any dissipation.
In this regime ~\cite{volkov1973collisionless, Barankov_2004}, 
after an external perturbation each pseudospin $\mathbf{s}_k$ starts to precess
with frequency $E_k = \sqrt{\xi_k^2 + |\Delta|^2}$. 
While there is no collision between quasiparticles,
the gap defined by the momentum sum $\Delta(t) = \sum_k U_{ij}\langle s_k^- \rangle$ 
can display algebraic decay through destructive interference of 
modes oscillating at different frequencies. This is an analog of
Landau damping~\cite{volkov1973collisionless}:
individual pseudospins continue to precess indefinitely, and in 
our finite-time numerical window this manifests as long-lived oscillations of $|\Delta_i(t)|$. 
In the 2D map, these oscillations fill the spectrum 
with the extended spectral weight along $\omega_{\Delta t}$, connecting all resonance features into 
continuous vertical stripes.

\begin{figure}[h!] 
    \centering 
 \includegraphics[scale=0.47]{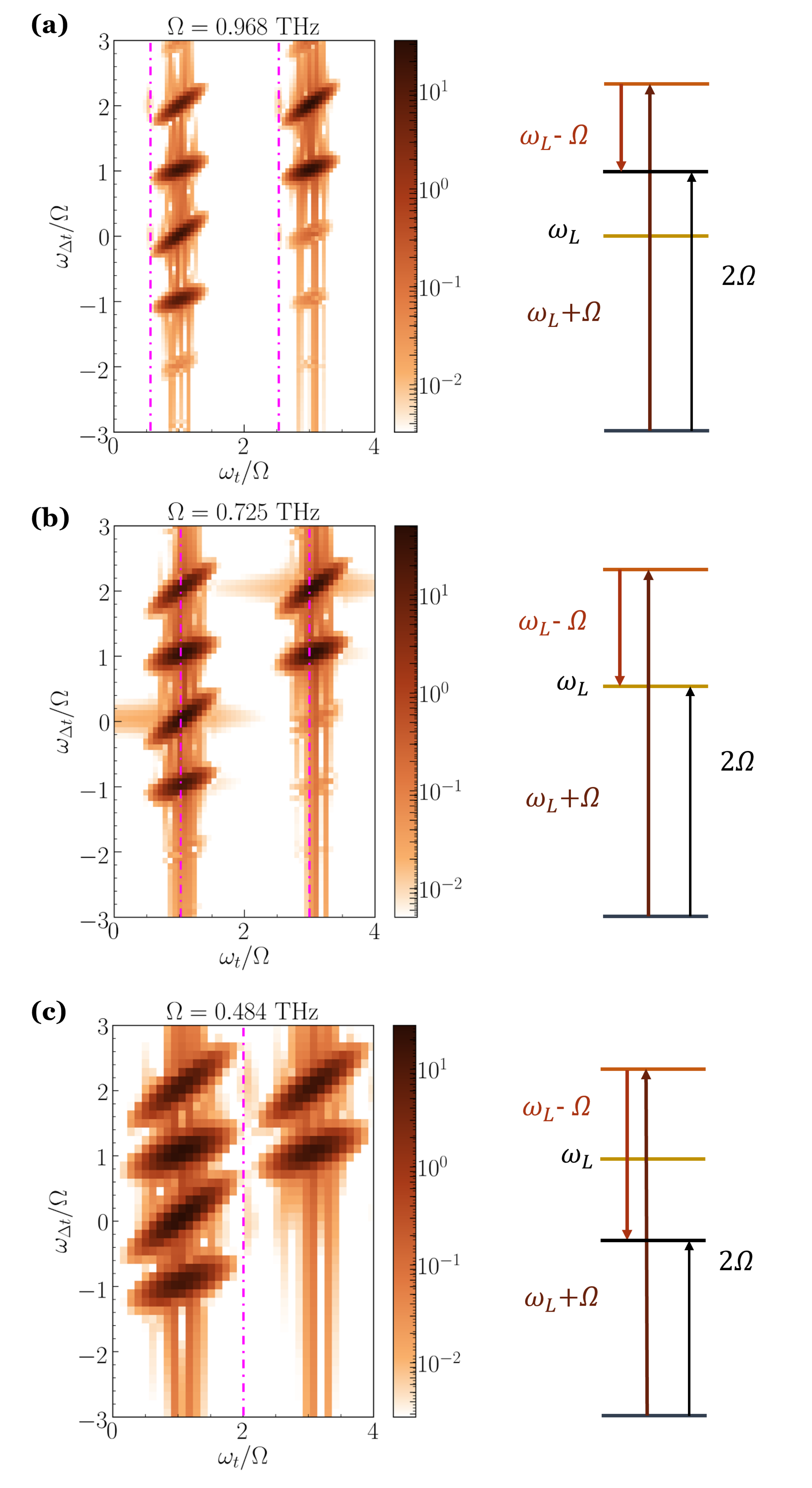} 
    \caption{
Frequency-sweep identification of the between-gap 
Leggett mode ($\omega_L \approx 6.2\,\mathrm{meV}$, $T = 4\,\mathrm{K}$, 
$\gamma = 0.1$~meV). Each row shows $|j_{\rm NL}(\omega_t, \omega_{\Delta t})|$ 
(logarithmic scale, left) and the corresponding energy level scheme 
(right) for carrier frequencies (a) $\Omega = 0.968\,\mathrm{THz}$, 
(b) $\Omega = 0.725\,\mathrm{THz}$, and (c) $\Omega = 0.484\,\mathrm{THz}$, 
with $\omega_L$ fixed. The magenta dash-dotted lines mark the predicted 
sideband positions $\omega_t =\pm \Omega \pm \omega_L$. As $\Omega$ 
decreases, the sidebands sweep across the map predictably: in (a), 
where $2\Omega > \omega_L$, 
the upconverted and downconverted sidebands 
at $\pm\Omega \pm \omega_L$ appear as distinct vertical features; in (b), $2\Omega \approx \omega_L$ and the 
sidebands correspond in position with the first- and third-harmonic signals; in (c), 
$3\Omega \approx \omega_L$ and only the upconverted sideband is 
in the displayed window, approaching $\omega_t \approx -\Omega+\omega_L\approx 2\Omega$, with the other sideband is located at $\omega_t\approx \Omega+\omega_L \approx 4\Omega$. }
    \label{fig:final_0.32}
\end{figure}

With increasing $\gamma$ [Fig.~\ref{fig:relax_gap}(b) and (c)], the pseudospins are 
actively driven back toward equilibrium on a timescale $\sim 1/\gamma$, 
and the gap oscillations are exponentially damped. Accordingly, the 2D map changes 
in two ways: Along the delay axis $\omega_{\Delta t}$, since the Leggett-mode coherence 
excited by the pump decays on the timescale $1/\gamma$ before the 
probe arrives, contributions from large delays are exponentially 
attenuated, narrowing the extent of each feature in $\omega_{\Delta t}$. 
Along the emission axis $\omega_t$, as the long-lived oscillations are quenched, 
the width of the sidebands vary and the spectral weight is mostly centered at the hot spot.

Using the conversion $1\,\mathrm{a.u.}=10\,\mathrm{meV}$, the two
finite damping strengths considered here correspond to
$\gamma=0.1\,\mathrm{meV}$ and $0.5\,\mathrm{meV}$, yielding
relaxation timescales $\tau_\gamma=1/\gamma$ of approximately
$6.6\,\mathrm{ps}$ and $1.3\,\mathrm{ps}$, respectively.
Importantly, throughout the range of $\gamma$ considered, the Leggett-mode
sidebands at $\omega_t=\omega_L\pm\Omega$ remain clearly visible at
their expected spectral positions without any appreciable shift.
This demonstrates that the characteristic Leggett-mode signature is
robust against quasiparticle relaxation on experimentally relevant
timescales.

All simulations discussed above were performed for $\nu=0.02$, for
which the Leggett-mode frequency is $\omega_L\simeq3\,\mathrm{meV}$,
placing the mode below both quasiparticle thresholds.
In MgB$_2$, however, the Leggett mode is expected to lie between the two
gaps, $2\Delta_\pi\simeq5\,\mathrm{meV}$ and
$2\Delta_\sigma\simeq14\,\mathrm{meV}$.
In this between-gap regime, the Leggett mode can decay into
$\pi$-band quasiparticles and consequently acquires an intrinsic
linewidth through coupling to the quasiparticle continuum.

To examine how the characteristic 2DCS signatures evolve in this
regime, we perform simulations for $\nu=0.32$, as shown in
Fig.~\ref{fig:final_0.32}.
Compared to the in-gap case, the vertical sideband features become
weaker.
Importantly, their positions in the
$(\omega_t,\omega_{\Delta t})$ plane continue to encode the
Leggett-mode frequency.
The visibility of these features depends on the pulse parameters and
can be enhanced in the narrow-band regime.
A longer delay-time scan also improves the frequency resolution along
the $\omega_{\Delta t}$ direction.
These results indicate that the characteristic Leggett-mode signatures
can remain identifiable even in the experimentally relevant between-gap
regime.
A fully quantitative description of the intrinsic decay into the
quasiparticle continuum, however, goes beyond the present pseudospin
treatment.

\section{Conclusions}

In this work, we have theoretically demonstrated that THz
two-dimensional coherent spectroscopy provides characteristic
fingerprints of the Leggett mode in a two-band superconductor.
By numerically solving the Anderson pseudospin Bloch equations for a
model representative of MgB$_2$ driven by two time-delayed THz pulses,
we find that, in addition to the conventional harmonic peaks at
$\omega_t=\Omega$ and $3\Omega$, the 2D nonlinear spectrum exhibits
a pair of characteristic sidebands at
$\omega_t=\omega_L\pm\Omega$.
We independently verify that the frequency $\omega_L$ extracted from
these sideband positions coincides with the Leggett-mode frequency
obtained from the relative-phase oscillation
$\varphi_1(t)-\varphi_2(t)$ of the two condensates.

The physical origin of these sidebands can be understood as a two-stage
process.
First, the pump pulse nonlinearly drives the two-band system through the
$A^2(t)$ coupling, inducing a coherent relative-phase oscillation between
the two condensates, namely, the Leggett mode at frequency $\omega_L$.
This excitation is Raman-like in nature and is activated by the broad spectral content of 
the finite-width pulse, which provides the frequency components required
to excite the Leggett mode even when $\omega_L\neq\Omega$.
Second, the probe pulse mixes with the coherently excited Leggett mode,
generating sidebands at $\Omega\pm\omega_L$.
This nonlinear mixing process, and hence the resulting sideband signal,
is characteristic of the two-pulse 2DCS protocol and does not appear in
the corresponding single-pulse response.

On the analytical side, we have shown that the finite pulse duration is
not merely a practical detail but plays an essential physical role.
The Cauchy principal-value contributions to the pulse spectrum, which
are absent in the idealized monochromatic limit, provide the nonlinear
current with spectral weight away from $\pm\Omega$, thereby enabling
resonant access to the Leggett mode.
The resulting 2D spectrum is governed by two sets of resonance conditions
of the third-order susceptibility $\chi^{(3)}$, associated with the
$j_{A_1^2A_2}$ and $j_{A_1A_2^2}$ channels, respectively.
Observable hot spots emerge where these collective-mode resonance
conditions overlap with the spectral-weight structures imposed by the
pulse spectrum.

Two further results support the experimental relevance of our proposal.
First, the temperature evolution of the sideband positions follows
$\omega_L(T)$ as the superconducting gaps are suppressed upon approaching
$T_c$, providing an internal consistency check and a direct means of
tracking the Leggett-mode frequency as a function of temperature.
Second, including phenomenological Gilbert damping with
$\gamma=0.1$ to $0.5\,\mathrm{meV}$, corresponding to relaxation
timescales of approximately $6.6$ to $1.3\,\mathrm{ps}$, suppresses the
sideband intensity but leaves their spectral positions essentially
unchanged.
This demonstrates that the characteristic Leggett-mode signature remains
robust in the presence of experimentally relevant dissipation.

From an experimental perspective, our results are directly relevant to
the recent narrow-band THz-2DCS measurements on MgB$_2$ by Katsumi
\textit{et al.}~\cite{Katsumi2025}.
Although a clear Leggett-mode signature was not identified in those
measurements, our analysis suggests that its detection may be facilitated
by systematically varying the central driving frequency $\Omega$ as in Fig.\ref{fig:final_0.32}.
Because the Leggett-induced sidebands shift with $\Omega$ according to
the resonance conditions derived above, a frequency sweep provides a
way to distinguish them from other spectral features and to extract the
underlying mode frequency.
Narrower-band pulses and a longer delay-time scan may further enhance
the visibility and frequency resolution of these features.
These considerations suggest concrete extensions of existing THz-2DCS
experiments that could test the predicted Leggett-mode signatures.

More broadly, the present mechanism is not restricted to MgB$_2$ or to the Leggett mode.
More generally, it may provide a route to identifying Raman-active
collective excitations whose characteristic frequencies are separated
from the central driving frequency.
THz-2DCS may therefore offer a broadly applicable framework for
disentangling collective dynamics in superconductors, 
including iron-based superconductors and 
time-reversal symmetry broken superconductors with
in-gap collective modes.


\acknowledgements

We thank the Max Planck-UBC-UTokyo Center for Quantum Materials for fruitful collaborations and financial support. 
N.T. acknowledges support by JST FOREST (Grant No. JPMJFR2131) and JSPS KAKENHI (Grants Nos. JP24H00191, JP25H01246 and JP25H01251).

\bibliography{references.bib} 

\end{document}